\documentclass[superscriptaddress,secnumarabic,nobibnotes,aps,prd,showkeys,noshowpacs,onecolumn,nofootinbib]{revtex4-2}
\usepackage{graphics}
\usepackage{graphicx}
\usepackage{color}
\usepackage{epsf}
\usepackage{bm}
\usepackage{caption}
\usepackage{subcaption}
\usepackage{amsmath,amssymb,amsfonts,mathrsfs,amsthm}
\usepackage{latexsym}
\usepackage{enumerate}
\usepackage{comment}
\usepackage[dvipsnames,svgnames,x11names,table]{xcolor}
\usepackage[colorlinks = true,
            linkcolor = Blue,
            urlcolor  = Blue,
            citecolor = Blue,
            anchorcolor = NavyBlue]{hyperref}
\usepackage{epstopdf}

\newcommand{\chssk}[2]{\genfrac{\{}{\}}{0pt}{}{#1}{#2}}

\def\v{\mbox{\rm v}}

\newcommand{\be}{\begin{eqnarray}}
\newcommand{\ee}{\end{eqnarray}}
\newcommand{\ben}{\begin{equation*}}
\newcommand{\een}{\end{equation*}}
\newcommand{\bean}{\begin{eqnarray*}}
\newcommand{\eean}{\end{eqnarray*}}

\newcommand{\bsub}{\begin{subequations}}
\newcommand{\esub}{\end{subequations}}

\def\bal#1\eal{\begin{align}#1\end{align}}

\newcommand{\disfrac}[1][2]{\displaystyle\frac}

\newcommand{\non}{\nonumber}

\begin{document}

\title{Static, spherically symmetric solutions in $f(Q)$-gravity and in nonmetricity scalar-tensor theory}

\author{Nikolaos Dimakis}
\email{nikolaos.dimakis@ufrontera.cl}
\affiliation{Departamento de Ciencias F\'{\i}sicas, Universidad de la Frontera, Casilla
54-D, 4811186 Temuco, Chile}
\author{Petros A. Terzis}
\email{pterzis@phys.uoa.gr}
\affiliation{Nuclear and Particle Physics section, Physics Department, University of
Athens, 15771 Athens, Greece}
\author{Andronikos Paliathanasis}
\email{anpaliat@phys.uoa.gr}
\affiliation{Institute of Systems Science, Durban University of Technology, Durban 4000, South Africa}
\affiliation{Departamento de Matem\'{a}ticas, Universidad Cat\'{o}lica del Norte, Avda. Angamos 0610, Casilla 1280 Antofagasta, Chile}
\affiliation{School for Data Science and Computational Thinking, Stellenbosch University, 44 Banghoek Rd, Stellenbosch 7600, South Africa}
\affiliation{School of Technology, Woxsen University, Hyderabad 502345, Telangana, India}
\author{Theodosios Christodoulakis}
\email{tchris@phys.uoa.gr}
\affiliation{Nuclear and Particle Physics section, Physics Department, University of
Athens, 15771 Athens, Greece}

\begin{abstract}
We solve the gravitational field equations for a static, spherically symmetric spacetime within the framework of the symmetric teleparallel theory of gravity. Specifically, we derive new solutions within the context of power-law $f(Q)$ gravity and the nonmetricity scalar-tensor theory. For the connection in the non-coincidence gauge, we present the point-like Lagrangian that describes the employed field equations. Furthermore, we construct two conservation laws, and for different values of these conserved quantities, we analytically solve the gravitational field equations. New solutions are obtained, we investigate their physical properties and their general relativistic limit. Finally, we discuss the algebraic properties for the derived spacetimes.
\end{abstract}

\keywords{$f(Q) $-gravity; black holes; exact solutions;
nonmetricity scalar-tensor.}
\maketitle

\section{Introduction}

The separation of the connection from the metric tensor is a subject of debate in gravitational physics. If we assume a metric tensor along a general connection, then the latter can be decomposed into two parts: the symmetric term and the antisymmetric component. The symmetric term can be separated further into the Levi-Civita connection and the nonmetricity components. Each component of the decomposition of the connection leads to the construction of a tensor. The Levi-Civita component of the connection defines the curvature $R_{\;\lambda \mu \nu }^{\kappa }$, the antisymmetric component defines the torsion tensor $\mathrm{T}_{\mu \nu
}^{\lambda }$, and the nonmetricity tensor $Q_{\kappa \mu \nu }$  is defined by the remaining components of the connection \cite{eisenhart}.

In General Relativity (GR), the Ricci scalar derived by the curvature, defines the fundamental Lagrangian density of the gravitational theory. The torsion and the nonmetricity tensors can be used to construct scalars, and when the latter are utilized to define a gravitational theory, they lead to the Teleparallel Equivalent of General Relativity (TEGR) \cite{ein28,Hayashi79,Maluf:1994ji} and the Symmetric Teleparallel Equivalent of General Relativity (STEGR) \cite{Nester:1998mp}. These two theories are equivalent to GR in the sense that the induced field equations give rise to the same metric. The three aforementioned scalars constitute the geometric trinity of gravity. For more details, we refer the reader to the discussion in \cite{trinity}.

General Relativity is challenged on cosmological scales \cite{Teg,Kowal,Komatsu,suzuki11,Ade15,ade18,cco1}, thus modified alternative theories of gravity have been introduced to explain the observable phenomena \cite{clifton,df2,meff1,meff2,meff3,rev1b}. Recently, part of the literature has focused on STEGR and its generalizations, such as the $f\left( Q\right)$-theory \cite{Koivisto2,Koivisto3}, the nonmetricity scalar-tensor theory \cite{Baha1,sc1,sc2,sc3}, and others \cite{mf1,mf2,mf3,mf4}. $f\left( Q\right)$-gravity is an extension of STEGR where nonlinear terms of the nonmetricity scalar $Q$ are introduced in the gravitational Action Integral, such that the gravitational Lagrangian becomes a nonlinear function $f$ of $Q$ \cite{rev1}. On the other hand, in the nonmetricity scalar-tensor theory, it is assumed that a scalar field is nonminimally coupled to the Lagrangian of the STEGR theory \cite{Baha1,sc1,sc2,sc3}. These generalized theories do not have an equivalent description within the framework of generalized GR theories. This has opened new directions for the study of nonmetricity theories in cosmology \cite{re1,re2,re3,re4,re5,re6,re7,re8,re9,re10,re11,re12,re13,re14,re15,re16,re17,re18,re19} and astrophysics \cite{rr1,rr2,rr3,rr4,rr5,rr6}. Although $f\left( Q\right)$-gravity suffers from the appearance of ghosts or from strong coupling in cosmological perturbations \cite{ppr1,ppr2}, it is pragmatic to examine the effects of the connection in gravity. In \cite{ppr3}, a solution was discussed regarding the viability of nonmetricity theories, while in \cite{Laur2} it has been suggested that the stability depends on the choice of a connection.

In this study, we are interested in the construction of exact solutions in $f\left( Q\right)$ and nonmetricity scalar-tensor theories within the framework of a static and spherically symmetric spacetime. We examine the exact solutions and discuss the appearance of singularities. We recover the results of \cite{BHHeis}, and we extend the analysis to the case of nonmetricity scalar-tensor gravity. Furthermore, we derive a new static spherically symmetric solution in $f(Q)\sim Q^{1+\kappa}$ theory, which reduces to the Schwarzschild black hole when $\kappa=0$. Inspired by \cite{tcl}, we investigate the case when this $f\left( Q\right)$ theory slightly deviates from STEGR.

The structure of the paper is as follows: In Section 2, we introduce the basic properties and definitions of STEGR. We define the gravitational models under consideration, which include symmetric teleparallel $f\left( Q\right)$-gravity and nonmetricity scalar-tensor theory. Specifically, we introduce the Brans-Dicke analogue in nonmetricity theory. In Section 3, we consider the geometry of a static, spherically symmetric spacetime. We write down the symmetric, flat connection, which inherits the symmetries of the background space. The connection depends on two arbitrary parameters, and the assumption that the field equations in $f\left( Q\right)$-gravity are diagonal leads to two distinct gravitational models. For each model, we write down the minisuperspace description and the gravitational Lagrangian function, from which we deduce the second-order gravitational field equations. These field
equations form a dynamical system describing the evolution of four scalars, i.e the three scale factors of the metric tensor and the two scalar fields introduced by the connection. To constrain the free function of the theory, we impose the Noether symmetry condition so that the gravitational Action Integral admits certain variational symmetries. As a result, we obtain the conditions under which the gravitational field equations admit conservation laws \cite{nrev1}. When the aforementioned conserved quantities are used, we end up with the $f(Q)\sim Q^{1+\kappa }$ gravitational models which are supported by this symmetry analysis.

By using the derived conservation laws for the gravitational field equations, in Section 4, we present exact solutions. We recover some previous results and derive new solutions. In Section 5, we extend our discussion to the framework of nonmetricity scalar-tensor theory, specifically in the case of the nonmetricity Brans-Dicke model. All the exact solutions derived for the $f\left( Q\right)$-theory remain valid solutions in the case of nonmetricity scalar-tensor theory for the power-law potential function. We remark that with the introduction of a Lagrange multiplier, $f\left( Q\right)$-theory can be written in a specific form of the nonmetricity Brans-Dicke theory. Furthermore, in Section 6 we discuss the physical properties of the effective geometric fluid related to the analytic solutions. Finally, in Section 7, we draw our conclusions.

\section{Symmetric teleparallel gravity}

The basic elements of STEGR and its extensions, namely $f(Q)$-theory, and nonmetricity scalar-tensor theory, are briefly presented in the following lines; for further reading we refer the interested reader to \cite{rev1}.

\subsection{STEGR}

Consider the four-dimensional manifold $V^{4}$ induced with the metric
tensor $g_{\mu \nu }$ and the covariant derivative $\nabla $ defined by the
generic connection $\Gamma_{\;\mu \nu }^{\alpha }$.

The  $\Gamma_{\;\mu \nu }^{\alpha }$ can be decomposed as
follows
\begin{equation}
\Gamma_{\;\mu \nu }^{\alpha }=\chssk{\alpha }{\mu \nu } +K_{\;\mu \nu
}^{\alpha }+L_{\; \mu \nu }^{\alpha }  \label{generalconnection}
\end{equation}%
where $\chssk{\alpha }{\mu \nu }$ are the Christoffel symbols%
\begin{equation}
\chssk{\alpha }{\mu \nu }=\frac{1}{2}g^{\lambda \alpha }\left(
g_{\lambda \mu ,\nu }+g_{\nu \lambda ,\mu }-g_{\mu \nu ,\lambda }\right) ,
\end{equation}%
$K_{\phantom{\alpha}\mu \nu }^{\alpha }$ is the contorsion part
\begin{equation}
K_{~\mu \nu }^{\alpha }=\frac{1}{2}\left( T_{~\mu \nu }^{\alpha }+\ T_{\mu
~\nu }^{~a}+T_{\nu ~\mu }^{~a}\right) ~,
\end{equation}%
constructed by the torsion $T_{~\mu \nu }^{\alpha }$ tensor, and $L_{~\mu \nu
}^{\lambda }$ is the disformation tensor defined by the nonmetricity
\begin{equation}
Q_{\alpha \mu \nu }=\nabla _{\alpha }g_{\mu \nu },
\end{equation}
as follows
\begin{equation}
L_{~\mu \nu }^{\lambda }=\frac{1}{2}g^{\lambda \sigma }\left( Q_{\mu \nu
\sigma }+Q_{\nu \mu \sigma }-Q_{\sigma \mu \nu }\right)   \label{disften}
\end{equation}

The different parts of the general connection (\ref{generalconnection}) can be used to
define distinct geometric objects. Specifically, the Ricci scalar $R$ associated to the metric is constructed by the Levi-Civita
connection%
\begin{equation}
R=g^{\lambda \nu }g_{\kappa }^{\mu }R_{\;\lambda \mu \nu }^{\kappa
}=\chssk{\kappa  }{\lambda \nu }_{,\mu }-\chssk{\kappa  }{\lambda\mu}_{,\nu }+\chssk{\sigma  }{\lambda \nu }\chssk{\kappa  }{\mu\sigma}-\chssk{\sigma  }{\lambda\mu}\chssk{\kappa  }{\mu\sigma}.
\end{equation}
In addition to this we have the torsion scalar $T$, which is related to the contorsion part of the connection,
that is,
\begin{equation}
T=\frac{1}{2}({K^{\mu \nu }}_{\beta }+\delta _{\beta }^{\mu }{T^{\theta \nu }%
}_{\theta }-\delta _{\beta }^{\nu }{T^{\theta \mu }}_{\theta }){T^{\beta }}%
_{\mu \nu },
\end{equation}
and the nonmetricity scalar $Q$ is defined by the disformation tensor as
follows
\begin{equation}
Q=-g^{\mu \nu }\left( L_{~\beta \mu }^{\alpha }L_{~\nu \alpha }^{\beta
}-L_{~\beta \alpha }^{\alpha }L_{~\mu \nu }^{\beta }\right) .  \label{Qdef}
\end{equation}

In General Relativity, connection $\Gamma _{~\mu \nu }^{\alpha }$ is
identified with the Levi-Civita connection, that is,~$\Gamma _{~\mu \nu
}^{\alpha }=\chssk{\alpha  }{\mu\nu}~,~K_{~\mu \nu }^{\alpha
}=0~,~L_{~\mu \nu }^{\alpha }=0.$ The only nonzero geometric scalar is the
Ricci scalar $R$, which is the Lagrangian density of Einstein's
gravitational theory.

On the other hand, in teleparallelism, $\Gamma _{~\mu \nu }^{\alpha
}=K_{~\mu \nu }^{\alpha }$,  the torsion scalar $T$ is the
fundamental quantity of TEGR with both the curvature and the nonmetricity
vanishing.

In the theory of symmetric teleparallelism the connection is considered to
be flat and symmetric, which leads to the conditions $R=0$ and $T=0$. Thus, STEGR is
defined by the nonmetricity scalar $Q$, where the modified
Einstein-Hilbert Action reads
\begin{equation}
S_{Q}=\int \!\!d^{4}x\sqrt{-g}Q.  \label{Qaction}
\end{equation}

Variation with respect to the metric tensor of the Action Integral (\ref%
{Qaction}) leads to the gravitational field equations%
\begin{equation}
\left( P_{\mu \rho \sigma }Q_{\nu }^{\;\rho \sigma }-2Q_{\rho \sigma \mu
}P_{~\nu }^{\rho \sigma }\right) +\frac{2}{\sqrt{-g}}\nabla _{\lambda
}\left( \sqrt{-g}P_{\;\mu \nu }^{\lambda }\right) =0
\end{equation}%
in which
\begin{equation}
P_{\;\mu \nu }^{\lambda }=-\frac{1}{4}Q_{\;\mu \nu }^{\lambda }+\frac{1}{2}%
Q_{(\mu ~\nu )}^{~~\lambda }+\frac{1}{4}\left( Q^{\lambda }-\bar{Q}^{\lambda
}\right) g_{\mu \nu }-\frac{1}{4}\delta _{\;(\mu }^{\lambda }Q_{\nu )},
\label{defP}
\end{equation}%
and $Q_{\mu }=Q_{\mu \nu }^{~~~~\nu }$ and $\bar{Q}_{\mu }=Q_{~\mu \nu
}^{\nu }$.

At this point it is important to mention that the Ricci scalar associated with
the Levi-Civita connection of the metric tensor $g_{\mu \nu }$ is related to
the nonmetricity scalar $Q$ of the symmetric and flat connection as
\begin{equation}
Q=R-\frac{1}{2}\mathring{\nabla}_{\lambda }P^{\lambda }~,~P^{\lambda
}=P_{~\mu \nu }^{\lambda }g^{\mu \nu }
\end{equation}%
where $\mathring{\nabla}_{\lambda }$ remarks covariant derivative with
respect to the Levi-Civita connection.

By replacing in the Action Integral (\ref{Qaction}) it follows $S_{Q}\simeq
\int \!\!d^{4}x\sqrt{-g}R$ (equivalence up to a boundary term), which is the Einstein-Hilbert Action of General
Relativity.

\subsection{\texorpdfstring{$f\left( Q\right) $-gravity}{f(Q)-gravity}}

Symmetric teleparallel $f\left( Q\right) $-gravity is a generalization of
STEGR, where the Action Integral (\ref{Qaction}) is modified so as to
include nonlinear components of the nonmetricity scalar $Q$.

The modified gravitational Action assumes the form
\begin{equation}
S_{f\left( Q\right) }=\int d^{4}x\sqrt{-g}f(Q),  \label{fqa}
\end{equation}%
where $f\left( Q\right) $ is an arbitrary smooth function. When $f\left(
Q\right) $ is linear, i.e. $f^{\prime \prime }(Q)=0$, the STEGR is recovered.
In the following, we use the prime to denote differentiation with respect to
the argument of the function; in the above relation $f^{\prime }(Q)=\frac{df%
}{dQ}$.

The gravitational field equations which correspond to the Action
Integral (\ref{fqa}) are \cite{Hohmann,Heis2}
\begin{equation}
\frac{2}{\sqrt{-g}}\nabla _{\lambda }\left( \sqrt{-g}f^{\prime }(Q)P_{\;\mu
\nu }^{\lambda }\right) -\frac{1}{2}f(Q)g_{\mu \nu }+f^{\prime }(Q)\left(
P_{\mu \rho \sigma }Q_{\nu }^{\;\rho \sigma }-2Q_{\rho \sigma \mu
}P_{~~~~\nu }^{\rho \sigma }\right) =0,  \label{fl1}
\end{equation}%
Moreover, the equation of motion for the connection reads \cite{Hohmann,Heis2}
\begin{equation}
\nabla _{\mu }\nabla _{\nu }\left( \sqrt{-g}f^{\prime }(Q)P_{~~~\sigma
}^{\mu \nu }\right) =0.  \label{kl.01}
\end{equation}

Since the connection is flat, there can always be found a coordinate system in which its components $\Gamma_{\;\mu \nu }^{\alpha }$ are zero. This is referred to as the coincident gauge \cite{Koivisto2,Koivisto3}. Nevertheless, when we adopt a specific metric tensor, we introduce a proper coordinate system which may not be compatible with the connection being zero \cite{Zhao}. Typical examples of this case are metrics with spherical symmetry. Of course this does not mean that spherical symmetry is incompatible with the coincident gauge. It is the typical spherical coordinates, in which the metric is usually given, which are inconsistent with having $\Gamma_{\;\mu \nu }^{\alpha }=0$. To clarify this, if the starting point is the line element
\begin{equation}
ds^{2}=-a(r)^{2}dt^{2}+n(r)^{2}dr^{2}+b(r)^{2}\left( d\theta ^{2}+\sin^{2}\theta \,d\phi ^{2}\right) ,  \label{lineelgen}
\end{equation}
then we are obligated to not use the coincident gauge, or else we over-restrict the equations of motion. The metric \eqref{lineelgen} can be transformed in a coordinate system compatible with having $\Gamma_{\;\mu \nu }^{\alpha }=0$, but in those coordinates it attains an over-complicated form, having all of its off-diagonal components non-zero. The explicit expressions of metrics compatible with the coincident gauge under spherical symmetry can be found in \cite{Laur1}. Here, we choose to use the simple form \eqref{lineelgen} for the metric, and thus adopt a non-zero connection. The connections which are compatible with the field equations in these coordinates have been found in \cite{Hohmann2,BHHeis}. The procedure consists of requiring that the isometries of the spacetime are also symmetries of the connection. We shall return to this point later in the analysis, when discussing the spacetime metric. 

Returning to the general theory, the gravitational field equations (\ref{fl1}) can be written in the
equivalent form \cite{Zhao}
\begin{equation}
f^{\prime }\left( Q\right) G_{\mu \nu }-\frac{1}{2}g_{\mu \nu }\left(
f\left( Q\right) -f^{\prime }\left( Q\right) Q\right) +2f^{\prime \prime
}\left( Q\right) P_{~~\mu \nu }^{\lambda }\nabla _{\lambda }Q=0
\end{equation}%
or, in a more compact manner
\begin{equation}
G_{\mu \nu }=\kappa_{eff}T_{\mu \nu }^{f\left( Q\right) } ,  \label{kl.02}
\end{equation}%
where $G_{\mu \nu }$ is the Einstein tensor, which in STEGR is expressed as
\begin{equation}
G_{\mu \nu }=\left( P_{\mu \rho \sigma }Q_{\nu }^{\;\rho \sigma }-2Q_{\rho
\sigma \mu }P_{\phantom{\rho\sigma}\nu }^{\rho \sigma }\right) +\frac{2}{%
\sqrt{-g}}\nabla _{\lambda }\left( \sqrt{-g}P_{\;\mu \nu }^{\lambda }\right)
,
\end{equation}%
and the $T_{\mu \nu }^{f\left( Q\right) }$ is the effective energy momentum
tensor, which includes the geometric degrees of freedom related to the
nonlinear function $f\left( Q\right) $, i.e.
\begin{equation}
T_{\mu \nu }^{f\left( Q\right) }=\frac{1}{2}g_{\mu \nu }\left( f\left(
Q\right) -f^{\prime }\left( Q\right) Q\right) -2f^{\prime \prime }\left(
Q\right) P_{~~\mu \nu }^{\lambda }\nabla _{\lambda }Q
\end{equation}%
and%
\begin{equation}
\kappa_{eff}=\frac{1}{f^{\prime }\left( Q\right) }.
\end{equation}%
The later is the effective time-varying modified Einstein \textquotedblleft
constant\textquotedblright .

We observe that when $Q=Q_{0}$, and $f^{\prime }\left( Q_{0}\right) \neq 0$, where $Q_0\in \mathbb{R}$, the limit of STEGR is recovered, with the effective energy momentum tensor $%
T_{\mu \nu }^{f\left( Q\right) }$ becoming
\begin{equation*}
T_{\mu \nu }^{f\left( Q\right) }=\frac{1}{2}g_{\mu \nu }\left( f\left(
Q_{0}\right) -f^{\prime }\left( Q_{0}\right) Q_{0}\right) ,
\end{equation*}%
contributing as an effective cosmological constant.

\subsection{Nonmetricity scalar-tensor gravity}

Another generalization of the STEGR comes from the introduction of a scalar field $%
\varphi$ nonminimally coupled to the nonmetricity $Q$. In this nonmetricity scalar-tensor theory the
gravitational Action Integral is defined as \cite{Baha1,sc1}%
\begin{equation}
S_{ST\varphi }=\int d^{4}x\sqrt{-g}\left( \frac{F\left( \varphi \right) }{2}%
Q-\frac{\omega \left( \varphi \right) }{2}g^{\mu \nu }\varphi _{,\mu
}\varphi _{,\nu }-V\left( \varphi \right) \right) ,  \label{ai.01}
\end{equation}%
in which $F\left( \varphi \right) $ is the coupling function between the
scalar field and the gravitational scalar $Q$. The $V\left( \varphi \right) $ is
the scalar field potential. The function, $\omega \left( \phi \right) $ is not
essential and it can be eliminated by introducing a new scalar field $d\Phi =%
\sqrt{\omega \left( \varphi \right) }d\varphi $.

The gravitational field equations of the nonmetricity scalar-tensor theory
are%
\begin{equation}
F\left( \varphi \right) G_{\mu \nu }+2F_{,\phi }\varphi _{,\lambda }P_{~~\mu
\nu }^{\lambda }+g_{\mu \nu }V\left( \varphi \right) +\frac{\omega \left(
\varphi \right) }{2}\left( g_{\mu \nu }g^{\lambda \kappa }\varphi _{,\lambda
}\varphi _{,\kappa }-\varphi _{,\mu }\varphi _{,\nu }\right) =0,
\label{ai.001}
\end{equation}%
while the equations of motion for the connection and the scalar field read
\begin{equation}
\nabla _{\mu }\nabla _{\nu }\left( \sqrt{-g}F\left( \varphi \right)
P_{~~~\sigma }^{\mu \nu }\right) =0  \label{ai.002}
\end{equation}
\begin{equation}
\frac{\omega \left( \varphi \right) }{\sqrt{-g}}g^{\mu \nu }\partial _{\mu
}\left( \sqrt{-g}\partial _{\nu }\varphi \right) +\frac{\omega _{,\varphi }}{%
2}g^{\lambda \kappa }\varphi _{,\lambda }\varphi _{,\kappa }+\frac{1}{2}%
F_{,\varphi }Q-V_{,\varphi }=0.  \label{ai.003}
\end{equation}

In the limit where  $\omega \left( \varphi \right) =0$, $F\left( \varphi
\right) =\varphi $, the latter field equations take the functional form of $%
f\left( Q\right) $-theory \cite{Baha1,sc1}, in which $\varphi =f^{\prime }\left(
Q\right) $ and $V\left( \varphi \right) = f^{\prime }(Q)Q-f(Q) $. The expressions for $\varphi$ and $V(\varphi)$ are similar to those establishing the mapping between $f(R)$ gravity and the Brans-Dicke theory \cite{Tavakoli}, where in place of $Q$ there appears $R$.

\subsubsection{Brans-Dicke nonmetricity scalar-tensor gravity}

When $F\left( \varphi \right) =\varphi $ and $\omega \left( \varphi \right) =%
\frac{\omega }{\varphi },~\omega =$const., the gravitational Action Integral (%
\ref{ai.01}) becomes

\begin{equation}
S_{BD\varphi }=\int d^{4}x\sqrt{-g}\left( \frac{\varphi }{2}Q-\frac{\omega }{%
2\varphi }g^{\mu \nu }\varphi _{,\mu }\varphi _{,\nu }-V\left( \varphi
\right) \right) .
\end{equation}%
The parameter $\omega$ plays a similar role as that of the Brans-Dicke parameter \cite{BransDicke}
and this gravitational model can be seen as the analogue of the Brans-Dicke
theory is nonmetricity gravity.

By introducing the new field $\phi =\ln \varphi $, we can write the last
Action Integral in the equivalent form of the dilaton field, that is,
\begin{equation}
S_{D}=\int d^{4}x\sqrt{-g}e^{\phi }\left( \frac{Q}{2}-\frac{\omega }{2}%
g^{\mu \nu }\phi _{,\mu }\phi _{,\nu }-\hat{V}\left( \phi \right) \right) ~,~%
\hat{V}\left( \phi \right) =V\left( \phi \right) e^{-\phi }.  \label{sd.01}
\end{equation}%
where the field equations read
\begin{equation}
G_{\mu \nu }+2\phi _{,\lambda }P_{~~\mu \nu }^{\lambda }+g_{\mu \nu }\hat{V}%
\left( \phi \right) +\frac{\omega }{2}\left( g_{\mu \nu }g^{\lambda \kappa
}\phi _{,\lambda }\phi _{,\kappa }-\phi _{,\mu }\phi _{,\nu }\right) =0,
\end{equation}%
\begin{equation}
\frac{\omega }{\sqrt{-g}}g^{\mu \nu }\partial _{\mu }\left( \sqrt{-g}%
\partial _{\nu }\phi \right) +\frac{\omega }{2}g^{\lambda \kappa }\phi
_{,\lambda }\phi _{,\kappa }+\frac{1}{2}Q-\hat{V}_{,\varphi }=0
\end{equation}%
\begin{equation}
\nabla _{\mu }\nabla _{\nu }\left( \sqrt{-g}e^{\phi }P_{~~~\sigma }^{\mu \nu
}\right) =0.
\end{equation}

Due to the extra degrees of freedom offered by the connection, the theory is not exactly equivalent to the usual Brans-Dicke theory introduced as a modification to GR. This, is expected to imply less stringent limits on the Brans-Dicke parameter $\omega$ induced by solar system tests; for experimental constraints on the typical Brans-Dicke theory see \cite{testBD1,testBD2,testBD3,testBD4}.

\section{\texorpdfstring{$f(Q)$ static spherically symmetric spacetime}{f(Q) static spherically symmetric spacetime}}

We now consider a static and spherically symmetric spacetime with the line element given by \eqref{lineelgen}. The connections which share the symmetries of this line element have been calculated in \cite{Hohmann2,BHHeis}. To briefly discuss the process: each of the isometries $\xi$ of \eqref{lineelgen} are used to require $\mathcal{L}_\xi \Gamma^{\alpha}_{\;\;\mu\nu}=0$, where $\mathcal{L}$ signifies the Lie derivative. In addition of course $\Gamma^{\alpha}_{\;\;\mu\nu}$ needs to be a symmetric and flat connection, so that there is no torsion and no curvature. With this process, two general families of connections are derived. In this work we are going to consider just one of them. The reason is that the connection which we do not utilize here, leads to metric field equations with a non-vanishing non-diagonal component, which imposes the theory to be either linear in $Q$ or have $Q=$const. Both of these cases lead to dynamics equivalent to those of GR with or without a cosmological constant, which is a case well studied. A way to avoid this restriction over the theory or the nonmetricity scalar, would be to consider an appropriate matter content leading to an off-diagonal energy momentum tensor. We restrict however our analysis to vacuum solutions and thus proceed by considering the other family of connections where the off-diagonal terms can be avoided in vacuum without setting necessarily $f(Q)\sim Q$ or $Q=$const. 

The aforementioned connection, which we use here, has the following non-zero components
\begin{equation}
\begin{split}
& \Gamma _{\;tt}^{t}=c_{1}+c_{2}-c_{1}c_{2}\gamma _{1},\quad \Gamma
_{\;tr}^{t}=\frac{c_{2}\gamma _{1}(c_{1}\gamma _{1}-1)}{\gamma _{2}},\quad
\Gamma _{\;rr}^{t}=\frac{\gamma _{1}(1-c_{1}\gamma _{1})(c_{2}\gamma
_{1}+1)-\gamma _{2}\dot{\gamma}_{1}}{\gamma _{2}^{2}}, \\
& \Gamma _{\;\theta \theta }^{t}=-\gamma _{1},\quad \Gamma _{\phi \phi
}^{t}=-\sin ^{2}\theta \gamma _{1},\quad \Gamma
_{\;tt}^{r}=-c_{1}c_{2}\gamma _{2},\quad \Gamma _{\;tr}^{r}=c_{1}\left(
c_{2}\gamma _{1}+1\right) , \\
& \Gamma _{\;rr}^{r}=\frac{1-c_{1}\gamma _{1}(c_{2}\gamma _{1}+2)-\dot{\gamma%
}_{2}}{\gamma _{2}},\quad \Gamma _{\;\theta \theta }^{r}=-\gamma _{2},\quad
\Gamma _{\;\phi \phi }^{r}=-\gamma _{2}\sin ^{2}\theta ,\quad \Gamma
_{\;t\theta }^{\theta }=c_{1}, \\
& \Gamma _{\;r\theta }^{\theta }=\frac{1-c_{1}\gamma _{1}}{\gamma _{2}}%
,\quad \Gamma _{\phi \phi }^{\theta }=-\sin \theta \cos \theta ,\quad \Gamma
_{t\phi }^{\phi }=c_{1},\quad \Gamma _{\;r\phi }^{\phi }=\frac{1-c_{1}\gamma
_{1}}{\gamma _{2}},\quad \Gamma _{\;\theta \phi }^{\phi }=\cot \theta  .
\end{split}
\label{consol1}
\end{equation}%
The parameters $c_{1}$, $c_{2}$ are constants and $\gamma _{1}$, $\gamma _{2}$
are functions of $r$. The dot hereafter denotes differentiation with respect to $r$.

When we calculate the field equations, we see that also for this connection there exist a
non-diagonal field equation which is
\begin{equation}
\left( c_{1}+\frac{c_{2}}{2}-c_{1}c_{2}\gamma _{1}\right) \dot{Q}f^{\prime
\prime }(Q)=0,
\end{equation}%
But in this case it is not necessary to set $f(Q)\sim Q$ or $Q=$const. We can avoid the known GR solutions by selecting either
\begin{equation}
\gamma _{1}=\frac{2c_{1}+c_{2}}{2c_{1}c_{2}},  \label{set1}
\end{equation}%
or
\begin{equation}
c_{1}=c_{2}=0,\quad \text{with}\;\gamma _{1}\;\text{arbitrary}.  \label{set2}
\end{equation}
We can thus study deviations from the usual GR solutions.

\subsection{Minisuperspace Lagrangians}

In \cite{fqmin} it has been shown that the gravitational field equations of $%
f\left( Q\right) -$gravity in the case of cosmological spacetimes admit a minisuperspace description. The
existence of a minisuperspace is essential for the analysis of the
gravitational model. The method of variational symmetries (i.e.\
Noether symmetry analysis) can be easily applied in these reduced systems for the derivation of conservation
laws as in the case of classical mechanics. The emanating conserved quantities can be later used for the construction of exact solutions and for
the investigation of integrability properties of the field equations. In addition to the above, a minisuperspace description is essential for the investigation of quantum properties through quantum cosmology, which is supposed to be a the limit of a more general quantum gravitational theory.

For the static spherically symmetric model of our consideration for each
case of the two choices \eqref{set1}, \eqref{set2} we write the following minisuperspace Lagrangians
\begin{equation}
\begin{split}
L_{1}\left( n,a,\dot{a},b,\dot{b},Q,\dot{Q},\psi ,\dot{\psi}\right) =& \frac{%
1}{n}\left( 2f^{\prime }(Q)\left( 2b\dot{a}\dot{b}+a\dot{b}^{2}\right) -%
\frac{ab^{2}(c_{2}-2c_{1})^{2}}{4c_{1}c_{2}}f^{\prime \prime }(Q)\dot{Q}\dot{%
\psi}\right)  \\
& +n\left[ \left( 2a-\frac{c_{1}c_{2}b^{2}}{a}\right) f^{\prime \prime }(Q)%
\frac{\dot{Q}}{\dot{\psi}}+a\left( \left( 2-b^{2}Q\right) f^{\prime
}(Q)+b^{2}f(Q)\right) \right] ,
\end{split}
\label{minlag1}
\end{equation}%
and
\begin{equation}
\begin{split}
L_{2}\left( n,a,\dot{a},b,\dot{b},Q,\dot{Q},\psi ,\dot{\psi}\right) =& \frac{%
1}{n}\left( 2f^{\prime }(Q)\left( 2b\dot{a}\dot{b}+a\dot{b}^{2}\right)
+2ab^{2}f^{\prime \prime }(Q)\dot{Q}\dot{\psi}\right)  \\
& +n\left[ 2af^{\prime \prime }(Q)\frac{\dot{Q}}{\dot{\psi}}+a\left( \left(
2-b^{2}Q\right) f^{\prime }(Q)+b^{2}f(Q)\right) \right]
\end{split}
\label{minlag2}
\end{equation}%
respectively, where we have introduced the scalar field $\dot{\psi}$ such
that $\gamma _{2}(r)=1/\dot{\psi}$. Note that in this setting, and taking in account the transformation law of the connection $\Gamma^{\mu}_{\;\kappa\lambda}$ with components \eqref{consol1}, the $\psi(r)$ transforms as a scalar under diffeomorphisms $r=r(\bar{r})$.

\subsection{Variational symmetries}

The Noether symmetry analysis is a powerful approach in order to determine
variational symmetries and conservation laws. It has been widely applied in the study of gravitational physics and
cosmology; for a review we refer the reader in \cite{nrev1}. Several applications of
Noether's theorems in modified theories of gravity can be found in \cite%
{no1,no2,no3,no4,no5,no6,no7}.

Specifically, Noether's first theorem has been applied in order to
constraint the free functions and parameters of a given gravitational
theory, in order for the Action Integral to admit variational symmetries; for
more details we refer the reader to the discussion in \cite{nrev1}. According
to Noether's first theorem, the existence of a variational symmetry implies the emergence of a conservation law \cite{nrev1}. The
latter can be used to reduce the gravitational field equations.

The Lagrangian functions (\ref{minlag1}),~(\ref{minlag2}) describe constrained
dynamical systems. The Noether symmetry analysis for constrained dynamical
systems in cosmology has been widely discussed before in \cite{ndim1,ndim2}. In the
following lines we omit the presentation for the basic elements of the
symmetry analysis;  we present directly the main results.

The dynamical variables of the gravitational equations are the scale factors
$\left\{ n,a,b\right\} $, the nonmetricity scalar $Q$, and the scalar field $%
\psi $ related to the connection. Therefore we consider a generator of the
form
\begin{equation}
X=\xi _{0}(n,a,b,Q,\psi )\frac{\partial }{\partial n}+\xi _{1}(n,a,b,Q,\psi )%
\frac{\partial }{\partial a}+\xi _{2}(n,a,b,Q,\psi )\frac{\partial }{%
\partial b}+\xi _{3}(n,a,b,Q,\psi )\frac{\partial }{\partial Q}+\xi
_{4}(n,a,b,Q,\psi )\frac{\partial }{\partial \psi }
\end{equation}%
to describe an infinitesimal transformation in the five dimensional space of
the dynamical variables.

We obtain the infinite dimensional symmetry of the parametrization
invariance
\begin{equation}
X_{par}=\chi (r)\frac{\partial }{\partial r}-\frac{n}{2}\dot{\chi}(r)\frac{%
\partial }{\partial n},
\end{equation}%
with $\chi (r)$ an arbitrary function. According to Noether's second theorem such an infinite dimensional symmetry implies the existence of a constraint. This is the equivalent of the
time parametrization invariance in cosmological minisuperspace Lagrangians,
where in place of $n$ we have the lapse function. Here, in the static,
spherically symmetric case, this role is played by the $\sqrt{g_{rr}}$
component of the metric, indicating the freedom of arbitrarily choosing the
radial variable.

Apart from the previous symmetry, which is not relevant for the construction
of conserved charges, we additionally obtain the symmetry
\begin{equation}
X_{1}=\frac{\partial }{\partial \psi } ,  \label{sym1}
\end{equation}%
for both Lagrangians and for an arbitrary $f(Q)$. This is expected since
there is no dependence of $\psi $ inside the Lagrangians.

Finally, we attain the scaling vector
\begin{equation}
X_{2}=n\frac{\partial }{\partial n}+\left( 2\kappa -1\right) a\frac{\partial
}{\partial a}+b\frac{\partial }{\partial b}-2Q\frac{\partial }{\partial Q},
\label{sym2}
\end{equation}%
which is a symmetry of $L_{1}$ only in the case $\kappa =1$ and under the
condition that $f(Q)\sim Q^{2}$, and a symmetry of $L_{2}$ for any $\kappa
\in \mathbb{R}$ and for a theory $f(Q)\sim Q^{1+\kappa }$.

The linear case $f(Q)\sim Q$ just returns the same symmetry generators as in
the case of general relativity \cite{Schwtchris}
\begin{equation}
X_{GR_{1}}=n\frac{\partial }{\partial n}+a\frac{\partial }{\partial a}+b%
\frac{\partial }{\partial b},\quad X_{GR_{2}}=\frac{n}{2b}\frac{\partial }{%
\partial n}-\frac{a}{2b}\frac{\partial }{\partial a}+\frac{\partial }{%
\partial b},\quad X_{GR_{3}}=\frac{n}{a^{2}b}\frac{\partial }{\partial n}-%
\frac{1}{ab}\frac{\partial }{\partial a}.
\end{equation}%
Comparing to the non-linear $f(Q)$ case, we observe that only the first, the
scaling symmetry survives in the case of $f(Q)\sim Q^{1+\kappa }$ or $%
f(Q)\sim Q^{2}$ depending on the choice of the connection.

\section{\texorpdfstring{Exact solutions of power law $f\left( Q\right) =f_{0}Q^{1+\protect%
\kappa }$-gravity}{Exact solutions of power law f(Q) gravity}}

From the two Lagrangians, let us consider the $L_{2}$ of \eqref{minlag2} for a theory of the form $f(Q)=f_{0}Q^{1+\kappa }$, which can recover the STEGR at some limit ($\kappa\rightarrow 0$), and which allows us to utilize both symmetries \eqref{sym1} and \eqref{sym2}.

The Euler-Lagrange equations for $(n,a,b,Q,\psi )$ are equivalent to:
\begin{equation}
\left[ 2a\left( 1-\frac{\dot{b}^{2}}{n^{2}}\right) -\frac{4b\dot{a}\dot{b}}{%
n^{2}}\right] f^{\prime }(Q)+2a\dot{Q}\left( \frac{1}{\dot{\psi}}-\frac{b^{2}%
\dot{\psi}}{n^{2}}\right) f^{\prime \prime}(Q) +a b^2\left( f(Q)-Qf^{\prime
}(Q)\right) =0  \label{euln}
\end{equation}%
\begin{equation}
\frac{2\dot{Q}}{n^{2}}\left( b^{2}\dot{\psi}-2b\dot{b}+\frac{n^{2}}{\dot{\psi%
}}\right) f^{\prime \prime }(Q)-\left( \frac{4b\dot{b}\dot{n}}{n^{3}}-\frac{2%
}{n^{2}}\left( 2b\ddot{b}+\dot{b}^{2}\right) \right) f^{\prime }(Q)-\left(
2-b^{2}Q\right) f^{\prime}(Q) - b^2f(Q)=0  \label{eula}
\end{equation}%
\begin{equation}
\frac{4\dot{Q}}{n^{2}}\left[ b\left( \dot{a}-a\dot{\psi}\right) +a\dot{b}%
\right] f^{\prime \prime }(Q)-\left[ \frac{4\dot{n}}{n^{3}}\left( b\dot{a}+a%
\dot{b}\right) -\frac{4}{n^{2}}\left( b\ddot{a}+\dot{a}\dot{b}+a\ddot{b}%
\right) \right] f^{\prime }(Q)-2ab\left( f(Q)-Qf^{\prime }(Q)\right) =0
\label{eulb}
\end{equation}%
\begin{equation}
\frac{2\dot{\psi}^{2}}{n^{4}}\left[ -2b\dot{b}\left( \dot{a}-a\dot{\psi}%
\right) +b^{2}\left( \dot{a}\dot{\psi}+a\ddot{\psi}\right) -a\dot{b}^{2}%
\right] +\frac{1}{n^{2}}\left[ 2\dot{a}\dot{\psi}+a\left( \left(
b^{2}Q-2\right) \dot{\psi}^{2}-2\ddot{\psi}\right) \right] -\frac{2ab^{2}}{%
n^{5}}\dot{n}\dot{\psi}^{3}+\frac{2a}{n^{3}}\dot{n}\dot{\psi}=0  \label{eulQ}
\end{equation}%
\begin{equation}
\begin{split}
& 2a\dot{Q}^{2}\left( \frac{1}{\dot{\psi}^{2}}-\frac{b^{2}}{n^{2}}\right)
f^{\prime \prime \prime }(Q)+2\Bigg[\frac{1}{\dot{\psi}^{2}}\left( \dot{a}%
\dot{Q}+a\ddot{Q}-\frac{2a\dot{Q}\ddot{\psi}}{\dot{\psi}}\right) +a\dot{n}%
\dot{Q}\left( \frac{b^{2}}{n^{3}}+\frac{1}{n\dot{\psi}^{2}}\right)  \\
& -\frac{b}{n^{2}}\left( b\left( \dot{a}\dot{Q}+a\ddot{Q}\right) +2a\dot{b}%
\dot{Q}\right) \Bigg]=0 .
\end{split}
\label{eulpsi}
\end{equation}

The conserved charges, owed to the symmetries \eqref{sym1} and %
\eqref{sym2}, assume the form
\begin{equation}
I_{1}=2f_{0}\kappa (\kappa +1)aQ^{\kappa -1}\dot{Q}\left( \frac{b^{2}}{n}-%
\frac{n}{\dot{\psi}^{2}}\right)   \label{cons1}
\end{equation}%
and
\begin{equation}
I_{2}=4f_{0}(\kappa +1)Q^{\kappa }\left( \frac{b^{2}\dot{a}}{n}+\frac{%
2\kappa ab\dot{b}}{n}-\frac{\kappa ab^{2}\dot{\psi}}{n}-\frac{\kappa an}{%
\dot{\psi}}\right)   \label{cons2}
\end{equation}
respectively.

The first integral, which emanates from generator $X_{1}$ of \eqref{sym1},
which has no $\partial _{n}$ contribution, is conserved due to the second
order equations. The integral of motion $I_{2}$ is conserved additionally by
virtue of the constraint, i.e. the first order equation \eqref{euln}, and
always for $f(Q)=f_{0}Q^{1+\kappa }$.

\subsection{\texorpdfstring{Solution $I_1=0$, $I_2\neq 0$}{Solution I1=0, I2 not zero}}

If the on mass shell constant value of the charge $I_1$ of \eqref{cons1} is zero, it
immediately implies (leaving aside the well studied case of $Q=$const.)
\begin{equation}  \label{soln1}
n^2 = b^2 \dot{\psi}^2.
\end{equation}
With the use of this expression in \eqref{cons2} we get
\begin{equation}
\dot{a} = 2 \kappa a \left(\dot{\psi}-\frac{\dot{b}}{b}\right)+\frac{I_2
\dot{\psi}}{4 f_0 (\kappa +1)b Q^{\kappa } }.
\end{equation}
Substitution of the above into the constraint equation \eqref{euln}, allows
us to obtain algebraically the dependence of $a$ with respect to the other
functions
\begin{equation}  \label{sola1}
a = \frac{I_2 \dot{b} }{f_0 Q^{\kappa } \left[2 (\kappa +1) (4 \kappa -1)
r^2 \dot{b}^2-8 \kappa (\kappa +1) r b \dot{b}-\kappa b^4 Q+2 (\kappa +1) b^2%
\right]},
\end{equation}
where we see why we need to consider $I_2\neq 0$. The $I_2=0$ case will be
treated separately.

The system \eqref{euln}-\eqref{eulpsi} consists of four second order
equations and the first order constraint \eqref{euln}, which means that only
three of the second order equations are independent. We thus have the
freedom of fixing the gauge by choosing one of the involved functions $n$, $a
$, $b$, $Q$, $\psi $ to be some function of $r$. In this case, it is
particularly convenient to set
\begin{equation}
\psi =\ln r.  \label{gauge1}
\end{equation}%
Then, applying \eqref{soln1}, \eqref{sola1} and \eqref{gauge1} into \eqref{eula} allows us
to obtain
\begin{equation}
\ddot{b}=\frac{\dot{b}^{2}}{2b}-\dot{b}\left( \kappa \frac{\dot{Q}}{Q}+\frac{%
1}{r}\right) +\frac{b}{2}\left( \frac{2\kappa \dot{Q}}{rQ}+\frac{1}{r^{2}}%
\right) -\frac{\kappa b^{3}Q}{4(\kappa +1)r^{2}}.  \label{diffb1}
\end{equation}%
Consecutive use of the above together with \eqref{soln1}, \eqref{sola1} and %
\eqref{gauge1} inside \eqref{euln}, turns the latter into a first order
relation for $Q$, which can be integrated to give
\begin{equation}
Q=\frac{2(\kappa +1)\left( b-r\dot{b}\right) ^{2}}{b^{3}\left( \kappa
b-(2\kappa +1)r\dot{b}\right) }.  \label{solQ1}
\end{equation}%
With this last relation, equation \eqref{diffb1} turns into
\begin{equation}
\ddot{b}=\frac{\left( b-r\dot{b}\right) \left( \left( 4\kappa ^{2}+1\right)
b-(2\kappa +1)(6\kappa +1)r\dot{b}\right) }{2(\kappa +1)(2\kappa +1)r^{2}b},
\end{equation}%
whose general solution is
\begin{equation} \label{fsolb}
b=\alpha r\left(r^{\frac{2\kappa -1}{2\kappa +1}}- \beta\right) ^{\frac{%
2(\kappa +1)}{1-4\kappa }},
\end{equation}
for which of course we need to consider $\kappa\neq 1/4$. In the end of this section we treat with the special solution for $\kappa=1/4$. For the time being let us assume a generic $\kappa$. With the expression \eqref{fsolb} at hand, we can go backwards and calculate $Q$, $a$ and $n$ from %
\eqref{solQ1}, \eqref{sola1} and \eqref{soln1} respectively. We thus obtain
\begin{align} \label{fsolQ}
Q& =\frac{8\left( 2\kappa ^{2}+\kappa -1\right) ^{2}r^{-\frac{4}{2\kappa +1}%
}\left( r^{\frac{2\kappa -1}{2\kappa +1}}-\beta \right) ^{\frac{5}{4\kappa -1%
}}}{\alpha^2 (4\kappa -1)(2\kappa +1)^{2}\left(  \beta\left(1 -4 \kappa \right)
+r^{\frac{2\kappa -1}{2\kappa +1}}\right) },   \\
a& =\left( r^{\frac{2\kappa -1}{2\kappa +1}}-\beta \right) ^{-\frac{\kappa +1%
}{1-4\kappa }}\left( \beta(4\kappa-1) - r^{\frac{2\kappa -1}{2\kappa +1}}
\right) ^{\kappa +1} \label{fsola}, \\ \label{fsoln}
n& =\alpha \left( r^{\frac{2\kappa -1}{2\kappa +1}}-\beta \right) ^{\frac{%
2\left( \kappa +1\right) }{1-4\kappa }},
\end{align}%
which of course with $\psi =\ln r$, solve the Euler-Lagrange equations %
\eqref{euln}-\eqref{eulpsi} and subsequently the field equations for an $%
f(Q)=f_{0}Q^{1+\kappa }$ theory with the connection choice we have
discussed. We need to note that in the expression for $a$, \eqref{fsola}, we
have eliminated a multiplicative constant including $I_{2}$ by considering a
scaling in time transformation since $a^{2}$ is the coefficient in $dt^{2}$
in the line element \eqref{lineelgen}. We thus see that the numerical value
of $I_{2}$ is of no importance in this case; it matters only whether $I_{2}$
is zero (separate case) or not. This is also true for the value of the constant $\alpha$. It is easy to check that $\alpha$ can be eliminated by an appropriate scaling in $r$ and $t$ followed by a redefinition of the constant $\beta$. With a transformation of this kind, $\alpha$ appears only additively in the expression for $\psi$. However for the gravitational system, only $\dot{\psi}$ is relevant so the constant disappears altogether when considering the gravitational configuration. Nevertheless, the value of $\alpha$ is useful to decide the region of definition of certain solutions, depending on the value of $\kappa$. One can consider $\alpha$ as a complex number of modulus $1$ which serves to keep the scale factors real under the various possible choices of $\kappa$.

For simplicity we can also express the analytic solution as follows%
\begin{align}
a\left( \bar{b}\right) & =\left( 2\beta \left( 2\kappa -1 \right) \left( \bar{%
b}\right) ^{-\frac{1}{2\left( \kappa +1\right) }}-\frac{1}{\alpha}\left( \bar{b}\right) ^{-%
\frac{2\kappa }{ \kappa +1 }}\right) ^{\kappa +1}~, \\
n& =\bar{b}~,
\end{align}%
in which $\bar{b}=\frac{b\left( r\right) }{r}$.

Notice that by setting $\kappa=0$, which corresponds to the symmetric teleparallel equivalent of GR, we obtain the Schwarzschild solution
\begin{equation} \label{STEGRsol}
  a(r) = \frac{\beta  r+1}{\beta  r-1}, \quad n(r)= \alpha  \left(\frac{1}{r}-\beta \right)^2, \quad b(r)= \frac{ \alpha (\beta  r-1)^2}{r}.
\end{equation}
It may not be obvious in these coordinates, but if we perform a transformation that makes $b(r)$ the radial variable, i.e. induce a change $r\rightarrow \bar{r}$ for which
\begin{equation} \label{defnewr}
  b(r) = \bar{r} \Rightarrow r(\bar{r}) = \frac{\sqrt{\bar{r}} \sqrt{4 \alpha  \beta +\bar{r}}+2 \alpha \beta +\bar{r}}{2 \alpha \beta ^2},
\end{equation}
then we arrive at the more familiar expressions
\begin{equation} \label{Schw}
  a(\bar{r}) = \left(1+ \frac{4 \alpha \beta }{\bar{r} }\right)^{\frac{1}{2}}, \quad n(r)= \left(1+ \frac{4 \alpha \beta }{\bar{r} }\right)^{-\frac{1}{2}}, \quad b(\bar{r})= \bar{r}.
\end{equation}
This is the Schwarzschild solution with the mass being given by $M=-2\alpha \beta$. Note that in order to arrive at \eqref{Schw}, the $a$, $b$ transform as scalars, while the $n$, which is the square root of the $dr^2$ coefficient of the line element, transforms as a density $n(r) dr = n(\bar{r}) d\bar{r}$. Notice that for the linear theory, $f(Q)\sim Q$, the scale factors are obtained by simply setting with $\kappa=0$ in \eqref{fsolb}, \eqref{fsola} and \eqref{fsoln}. However, the corresponding $Q$ is not obtained from \eqref{fsolQ}, the $Q$ for the Schwarzschild solution is zero, while the field associated to the connection is arbitrary since the equation for the connection is trivialized. In various steps in order to arrive to our solution we have assumed $f^{\prime}(Q)\neq 0$ and $Q\neq$const. It is remarkable that even after this, at least the expressions for the scale factors of the linear theory can be recovered with a simple substitution $\kappa=0$.

Let us now investigate the case where $\kappa $ is very small, that is, $%
f\left( Q\right) =Q^{1+\varepsilon }= Q + \varepsilon Q\ln Q+ \mathcal{O}(\varepsilon^2)$, with $\varepsilon<<1$. In this way we shall investigate how the solution \eqref{fsolb}-\eqref{fsoln} is affected by small deviations from the GR equivalent. For
this theory the scale factors $a,n$ and $b$, up to the first order in $\varepsilon$, read%
\begin{equation}
a\left( r\right) \simeq \frac{\beta  r+1}{\beta  r-1} +\frac{8 \beta  r \ln r+(\beta  r-1) \left((\beta  r+1) \ln \left(-\frac{\beta  r+1}{r}\right)-4 \beta  r-5 (\beta  r+1) \ln \left(\frac{1}{r}-\beta \right)\right)}{(\beta  r-1)^2}\varepsilon ,
\end{equation}%
\begin{equation}
n\left( r\right) \simeq \alpha \left( \frac{1}{r}-\beta \right) ^{2}+ \frac{2 \alpha (\beta  r-1)}{r^2} \left(5 (\beta  r-1) \ln \left(\frac{1}{r}-\beta \right)-4 \ln r\right) \varepsilon ,
\end{equation}%
\begin{equation}
b\left( r\right) \simeq \alpha\, r\left( \frac{1}{r}-\beta \right) ^{2} + \frac{2 \alpha  (\beta  r-1)}{r} \left(5 (\beta  r-1) \ln \left(\frac{1}{r}-\beta \right)-4 \ln r\right) \varepsilon .
\end{equation}

We introduce once more the variable $\bar{r}$ defined by \eqref{defnewr}, which makes $b$ the radial variable up to zero order in $\varepsilon$. The the above relations now become%
\begin{equation}
\begin{split}
a\left( \bar{r}\right) = & \sqrt{1+\frac{4 \alpha  \beta }{\bar{r}}} + \Bigg[ \frac{8 \alpha  \beta  \ln \left(\frac{\sqrt{\bar{r} } \sqrt{4 \alpha  \beta +\bar{r} }+2 \alpha  \beta +\bar{r} }{2 \alpha  \beta ^2}\right)}{\bar{r} }-\frac{2 \sqrt{4 \alpha  \beta +\bar{r} }}{\sqrt{\bar{r} }}\\
& +\frac{\sqrt{4 \alpha  \beta +\bar{r} } \ln \left(\frac{\sqrt{\bar{r} } \sqrt{4 \alpha  \beta +\bar{r} }-4 \alpha  \beta -\bar{r} }{2 \alpha }\right)}{\sqrt{\bar{r} }}-\frac{5 \sqrt{4 \alpha  \beta +\bar{r} } \ln \left(\frac{\bar{r} -\sqrt{\bar{r} } \sqrt{4 \alpha  \beta +\bar{r} }}{2 \alpha }\right)}{\sqrt{\bar{r} }}-2 \Bigg] \varepsilon,
\end{split}
\end{equation}%
\begin{equation}
n\left( \bar{r}\right) =\frac{1}{\sqrt{1+\frac{4 \alpha  \beta }{\bar{r}}}} + \frac{4 \left(\sqrt{\bar{r} }-\sqrt{4 \alpha  \beta +\bar{r} }\right) \ln \left(\frac{\sqrt{\bar{r} } \sqrt{4 \alpha  \beta +\bar{r} }+2 \alpha  \beta +\bar{r} }{2 \alpha  \beta ^2}\right)+10 \sqrt{\bar{r} } \ln \left(\frac{\bar{r} -\sqrt{\bar{r} } \sqrt{4 \alpha  \beta +\bar{r} }}{2 \alpha }\right)}{\sqrt{4 \alpha  \beta +\bar{r} }}\varepsilon,
\end{equation}%
\begin{equation}
  \begin{split}
b\left( \bar{r}\right) = & \bar{r} + \Bigg[4 \bar{r}  \ln \left(\frac{\sqrt{\bar{r} } \sqrt{4 \alpha  \beta +\bar{r} }+2 \alpha  \beta +\bar{r} }{2 \alpha  \beta ^2}\right)-4 \sqrt{\bar{r} } \sqrt{4 \alpha  \beta +\bar{r} } \ln \left(\frac{\sqrt{\bar{r} } \sqrt{4 \alpha  \beta +\bar{r} }+2 \alpha  \beta +\bar{r} }{2 \alpha  \beta ^2}\right) \\
& +10 \bar{r}  \ln \left(\frac{\bar{r} -\sqrt{\bar{r} } \sqrt{4 \alpha  \beta +\bar{r} }}{2 \alpha }\right) \Bigg]\varepsilon.
\end{split}
\end{equation}%
from where the deviation from Schwarzschild solution becomes aparent.

We observe that in the limit where $\bar{r}\rightarrow \infty $, the asymptotic
solution leads to the Minkowski space
\begin{align}
a(\bar{r})  &\simeq 1 -4 \varepsilon  \left(1+\ln (-\beta )\right), \\
n(\bar{r})  &\simeq 1+ 10 \varepsilon  \ln (-\beta ), \\
b(\bar{r})  &\simeq \bar{r} \left(1+  10 \varepsilon  \ln (-\beta ) \right) .
\end{align}
We remind that the parameter $\beta$ is associated to the mass of the black hole,  $M=-2\alpha \beta$, which for a positive $\alpha$, or equivalently for $\alpha=1$ (since $\alpha$ is a non-essential constant), corresponds to having $\beta$ negative, which explains the minus in the above logarithms.  We thus establish that at the STEGR limit the spacetime is asymptotically Minkowski. In \cite{Laur1} it has been noted that only through connection \eqref{consol1} a reasonable Minkowski limit can be attained. 

It is not clear whether the general spacetime, induced by the solution \eqref{fsolb}-\eqref{fsoln}, i.e.
\begin{equation} \label{oursol}
  \begin{split}
  ds^2  = & - \left( r^{\frac{2\kappa -1}{2\kappa +1}}-\beta \right)^{-\frac{2(\kappa +1)}{1-4\kappa }}\left( \beta(4\kappa-1) - r^{\frac{2\kappa -1}{2\kappa +1}}
\right) ^{2(\kappa +1)} dt^2 + \alpha^2 \left( r^{\frac{2\kappa -1}{2\kappa +1}}-\beta \right) ^{\frac{4\left( \kappa +1\right) }{1-4\kappa }} dr^2 \\
    & + \alpha^2 r^2\left(r^{\frac{2\kappa -1}{2\kappa +1}}- \beta\right)^{\frac{4(\kappa +1)}{1-4\kappa }} \left( d\theta^2 + \sin^2\theta d\phi^2 \right),
  \end{split}
\end{equation}
supports black hole solutions outside the $\kappa=0$ limit of STEGR (and equivalently GR), where we saw that the Schwarzschild solution was recovered. At a first sight, turning to the nonmetricity scalar \eqref{fsolQ} or even the Ricci scalar of the curvature associated to the Levi-Civita connection, we may notice the existence of at least two singularities, depending on the values of the parameters. However the $r$ variable in which we obtained this solution cannot be considered as the typical radius variable of spherical coordinates. In order to transform the metric into such coordinates may even involve a complex transformation. This is actually true also for the transformation \eqref{defnewr}, which we performed to obtain the Schwarzschild metric. As can be seen, inside the horizon the $r(\bar{r})$ becomes complex. Performing complex transformations from one metric to another may affect the existence of singularities, horizons, etc. 

\begin{figure}[t]\centering
\begin{subfigure}{.5\linewidth}
  \centering
  \includegraphics[scale=0.65]{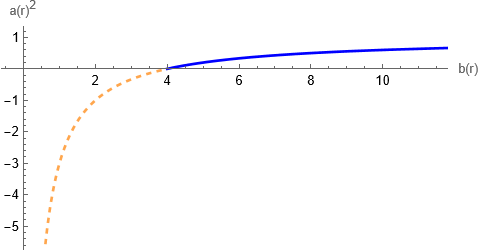}
  \caption{Parametric plot of $a(r)^2$ with respect to b(r).}
  \label{fig1:sub1}
\end{subfigure}%
\begin{subfigure}{.5\linewidth}
  \centering
  \includegraphics[scale=0.65]{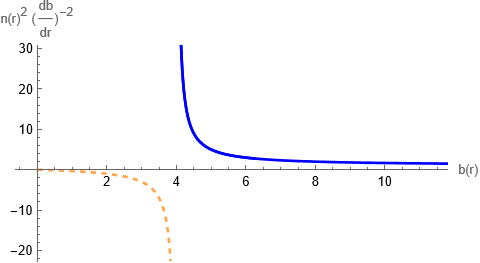}
  \caption{Parametric plot of $n(r)^2 \dot{b}^{-2}$ with respect to $b(r)$.}
  \label{fig1:sub2}
\end{subfigure}
\caption{The case of STEGR with $\kappa=0$, where the solution corresponds to the Schwarzschild metric. The continuous line corresponds to the parametric plots of $g_{tt}=a^2$ and $g_{\bar{r}\bar{r}}= n(r)^2 \dot{b}^{-2}$ with respect to $b(r)$ for $r\in\mathbb{R}$. The dashed line is the part that corresponds to the inside of the horizon, given by $g_{tt}(\bar{r})$, $g_{\bar{r}\bar{r}}(\bar{r})$ of \eqref{Schwmet} for $\bar{r}<2M=-4\alpha\beta$. The constants have been chosen as $\alpha=1$, $\beta=-1$.}
\label{fig1}
\end{figure}

To illustrate this situation, consider the $b(r)$, which acts as a radial coordinate in the line element \eqref{lineelgen}. From the solution \eqref{fsolb} it can be easily seen that the obtained $b(r)$ is not necessarily a monotonic function for $r\in\mathbb{R}$. Under appropriate values of the parameters there exists an extremum at
\begin{equation} \label{minr}
  r_0 = \left(4 \kappa ^2+1\right)^{\frac{2 \kappa +1}{1-2 \kappa }} (\beta  (4 \kappa -1) (2 \kappa +1))^{\frac{2 \kappa +1}{2 \kappa -1}} ,
\end{equation}
where $\dot{b}(r_0)=0$. For the STEGR with $\kappa=0$ this corresponds to $r_0=-1/\beta$, and it is a minimum with $b(r_0)= -4 \alpha \beta$. The latter is the distance where the horizon is placed, remember that in the Schwarzschild case $M=-2\alpha\beta$, hence $b(r_0)=2M$. This means that the line element
\begin{equation} \label{STEGRmet}
  ds^2 = - \left(\frac{\beta  r+1}{\beta  r-1}\right)^2 dt^2 + \alpha^2  \left(\frac{1}{r}-\beta \right)^4 dr^2 + \frac{ \alpha^2 (\beta  r-1)^4}{r^2} \left( d\theta^2+ \sin^2\theta d\phi^2 \right)
\end{equation}
constructed with \eqref{STEGRsol} cannot describe the full Schwarzschild spacetime for $r\in \mathbb{R}$, because in order for $b(r)<2M$, $r$ needs to be complex. The transformation, $r\rightarrow \bar{r}$, we performed in \eqref{defnewr} and which transforms \eqref{STEGRmet} to the Schwarzschild metric accomplishes this by being complex inside the horizon. It maps the $r\in \mathbb{C}$ values needed for $b(r)<2M=-4\alpha\beta$ to real values of the new variable $\bar{r}$, which is the actual radius of the Schwarzschild spacetime
\begin{equation} \label{Schwmet}
  ds^2 = - \left(1+ \frac{4 \alpha \beta }{\bar{r} }\right) dt^2 + \left(1+ \frac{4 \alpha \beta }{\bar{r} }\right)^{-1} dr^2 + \bar{r}^2 \left( d\theta^2+ \sin^2\theta d\phi^2 \right).
\end{equation}
In Fig. \ref{fig1} we depict the parametric graphs of $a^2$ and $n^2 \dot{b}^{-2}$ with respect to $b(r)$ for $\kappa=0$ (see relations \eqref{STEGRsol}). The $n^2 \dot{b}^{-2}$ corresponds to the coefficient of the $d\bar{r}^2$ part of the Schwarzschild line element in the coordinates of the radial variable $\bar{r}$, i.e. the $g_{\bar{r}\bar{r}}$. In the $r$ coordinate of metric \eqref{STEGRmet} the graphs are plotted considering $r\in\mathbb{R}$ (continuous line). The extended dashed line correspond to complex values of $r$ or equivalently to the inside of the horizon in the coordinates where the metric is given by \eqref{Schwmet}. In the $r$ variable description only the continuous line can be plotted for real values of the variables.

\begin{figure}[t]\centering
\begin{subfigure}{.5\linewidth}
  \centering
  \includegraphics[scale=0.65]{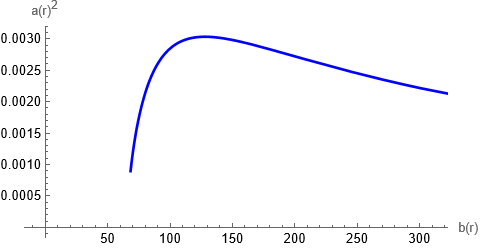}
  \caption{Parametric plot of $a(r)^2$ with respect to b(r).}
  \label{fig2:sub1}
\end{subfigure}%
\begin{subfigure}{.5\linewidth}
  \centering
  \includegraphics[scale=0.65]{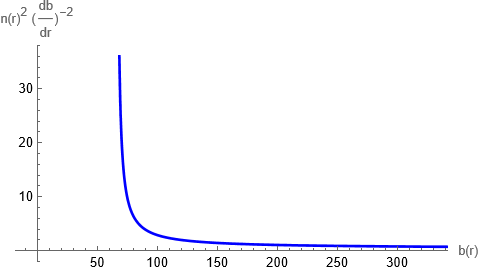}
  \caption{Parametric plot of $n(r)^2 \dot{b}^{-2}$ with respect to $b(r)$.}
  \label{fig2:sub2}
\end{subfigure}
\caption{Plots for $\kappa=1/6$, $\alpha=1$ and $\beta=-1$.}
\label{fig2}
\end{figure}

A similar, although not exactly identical, situation arises for other values of $\kappa$ as well. In Fig. \ref{fig2} we see the plots corresponding to $a^2$ and $n^2\dot{b}^{-2}$ as function of $b$ for the theory with $\kappa=1/6$. The graphs are drawn for a subset of the variable $r$ for which $b(r)$ is monotonic. For $r=r_0$ (see \eqref{minr}), the $b(r)$ has a minimum, $b_{\text{min}}=b(r_0)$. Values lesser than $b_{\text{min}}$ correspond to a complex $r$. It is not obvious if a complex transformation can be used assigning the region $(0,b_{\text{min}})$ to a real variable like it happens to the Schwarzschild case, leading to a real Lorentzian metric for all values of a positive radius.

\subsubsection{\texorpdfstring{The $\kappa=1/4$ case}{The kappa=1/4 case}}

The solution leading to \eqref{fsolb}-\eqref{fsoln} excludes the case $\kappa=1/4$, which we need to consider separately. For this particularly value of $\kappa$, the final expressions in the gauge $\psi = \ln r$ read
\begin{align}
  b & = b_1 r e^{-\frac{3 b_2}{r^{1/3}}} \\
  a & = \frac{e^{\frac{3 b_2}{2r^{1/3}}} \left(6 b_2+5 r^{\frac{1}{3}}\right)^{\frac{5}{4}}}{r^{\frac{5}{12}}}\\
  n & =  b_1 e^{-\frac{3 b_2}{r^{1/3}}}
\end{align}
with the nonmetricity scalar
\begin{equation} \label{fsolQsp}
  Q = -\frac{10 b_2^2 e^{\frac{6 b_2}{r^{1/3}}}}{b_1^2 r^{\frac{7}{3}} \left(6 b_2+5 r^{\frac{1}{3}}\right)} .
\end{equation}
The $b_1$, $b_2$ in the above expressions are non-zero constants of integration.

We can express the solution in the gauge where $b$ is the radial variable, say $b=\bar{r}$ by solving the relation
\begin{equation}
   b_1 r e^{-\frac{3 b_2}{r^{1/3}}} = \bar{r}
\end{equation}
and obtaining the transformation
\begin{equation}
  r(\bar{r}) = \frac{b_2^3}{W\left(x\right)^3} \, ,
\end{equation}
where $x = \frac{b_1^{1/3} b_2}{ \bar{r}^{1/3}}$ and $W(x)$ is the Lambert $W$ function, which is defined as the solution to $W e^W = x$. The $a$, $b$, $Q$ and $\psi$ transform as scalars while for $n$ we have, $n(r) dr = n(\bar{r}) d\bar(r)$, and thus we finally obtain
\begin{align}
  a & = \frac{\left(W\left(x\right)^{-1}+\frac{6}{5}\right)^{5/4}}{\sqrt{r} W\left(x\right)^{\frac{1}{4}}}  \\
  n & = \frac{1}{W\left(x\right)+1}\\
  Q & = -\frac{10 W\left(x\right)^2}{r^2 \left(6 W\left(x\right)+5\right)}\\
  \psi & = \ln \left( \frac{b_2^3}{W\left(x\right)^3} \right)\\
  b & = \bar{r}.
\end{align}
If both, $b_2$ and $b_1$ are positive, the corresponding spacetime has a singularity at $\bar{r}=0$, in the sense that the nonmetricity scalar, which is the basic geometric scalar of the theory, diverges at that point. The curvature scalars constructed with the curvature of the Levi-Civita connection, also diverge at the same point. An additional singularity can appear in a finite distance, if the constants are chosen so that the argument $x$ of the Lambert function is negative. This singularity appears at a radius for which $a\rightarrow 0$ (Killing horizon). Within that region the spacetime becomes complex with both $a$ and $n$ obtaining complex values. We thus have the following setting in what regards spacetime singularities: a) For $b_2>0$, which implies $x>0$, there exists one singularity at the origin $\bar{r}=0$, b) For $b_2<0$, i.e. $x<0$, there are two singularities, one at $\bar{r}=0$ and another at the finite nonzero distance corresponding to $a=0$.

\subsection{\texorpdfstring{Solution $I_1=0$, $I_2= 0$}{Solution I1=0, I2=0}}

In this case, apart from \eqref{soln1}, which also holds here due to setting
$I_1$ of \eqref{cons1} to zero, we additionally have
\begin{equation}  \label{sola2}
a = \frac{2 \kappa a}{b} \left(b \dot{\psi}-\dot{b}\right).
\end{equation}
With \eqref{soln1} and \eqref{sola2}, we can combine \eqref{euln} with %
\eqref{eulQ} to eliminate the quadratic term $\dot{b}^2$ and solve the
resulting equation algebraically with respect to $Q$, which yields
\begin{equation}  \label{solQ2}
Q = \frac{4 (\kappa +1) (2 \kappa -1) \left(\dot{b}-b \dot{\psi}\right)}{(2
\kappa +1) b^3 \dot{\psi}}.
\end{equation}
Then, use of this expression and \eqref{soln1}, \eqref{sola2} in \eqref{eulQ}
gives
\begin{equation}
\left(\dot{b}-b \dot{\psi}\right) \left((2 \kappa +1) (4 \kappa -1) \dot{b}%
-\left(4 \kappa ^2+1\right) b \dot{\psi}\right) =0.
\end{equation}
Satisfying the first parenthesis leads to $Q=0$, which we exclude since we
concentrate in $Q\neq$const. solutions that modify GR dynamics. Hence, the
above equation leads us to
\begin{equation}
b=\beta \exp \left(\frac{\left(4 \kappa ^2+1\right) \psi}{(2 \kappa +1) (4
\kappa -1)}\right) .
\end{equation}
Returning to the previous relations for the rest of the variables we obtain
\begin{subequations}
\label{solzeros}
\begin{align} \label{fsolQ00}
Q & = -\frac{8 (\kappa +1)^2 (2 \kappa -1)^2 }{(4 \kappa -1) (2 \beta \kappa
+\beta )^2} \exp \left[-\frac{2 \left(4 \kappa ^2+1\right) \psi}{(2 \kappa
+1) (4 \kappa -1)}\right] \\
a & = \exp \left[\frac{4 \kappa (\kappa +1) (2 \kappa -1) \psi}{(2 \kappa
+1) (4 \kappa -1)}\right] \\
n & = \beta \dot{\psi} \exp \left[\frac{\left(4 \kappa ^2+1\right) \psi}{(2
\kappa +1) (4 \kappa -1)}\right],
\end{align}
where $\psi$ remains an arbitrary function of $r$. This happened because we
did not perform a gauge fixing choice, as we did in the previous case, in
order to solve the differential equations. If we choose the gauge in which $b
$ becomes the radial variable, that is set
\end{subequations}
\begin{equation}
\beta \exp \left(\frac{\left(4 \kappa ^2+1\right) \psi}{(2 \kappa +1) (4
\kappa -1)}\right) = r \Rightarrow \psi = \frac{(2 \kappa +1) (4 \kappa -1)
}{4 \kappa ^2+1} \ln \left(\frac{r}{\beta }\right),
\end{equation}
the expressions \eqref{solzeros} become
\begin{align}
Q & = -\frac{8 \left(\kappa+1\right)^2\left(2\kappa-1\right)^2}{(4 \kappa
-1) (2 \kappa +1)^2 r^2} \\
a & = r^{\frac{4 \kappa (\kappa +1) (2 \kappa -1)}{4 \kappa ^2+1}} \\
n & = \frac{(2 \kappa +1) (4 \kappa -1)}{4 \kappa ^2+1},
\end{align}
where we once more exploited the fact that we can scale the time variable to
eliminate any constant appearing multiplicatively in the expression for $a(r)
$.

The line element in this case reads
\begin{equation}
ds^2 = - r^{\frac{8 \kappa (\kappa +1) (2 \kappa -1)}{4 \kappa ^2+1}} dt^2 +
\frac{(2 \kappa +1)^2 (4 \kappa -1)^2}{(4 \kappa ^2+1)^2} dr^2 + r^2 \left(
d\theta^2+ \sin^2\theta d\phi^2 \right) .
\end{equation}
This is a spacetime with a naked singularity at $r=0$. The nonmetricity scalar $Q$ diverges at $r=0$, and the same is true for the Ricci scalar corresponding to the Levi-Civita connection.

\subsection{\texorpdfstring{The $I_1\neq0$ and $I_2\neq 0$ case}{The I1 and I2 not zero case}}

The generic case where both $I_{1}$ and $I_{2}$ can be taken to be non-zero
is quite complicated. However, a special solution can be extracted. We
refrain from describing the process since it is a known solution that has
been reported before in the literature \cite{BHHeis}. The needed functions
are given, in our conventions, by
\begin{subequations} \label{Heissol}
\begin{align} \label{heisQ}
Q& =-\frac{8(\kappa +1)^{2}(2\kappa -1)^{2}}{(4\kappa -1)(2\kappa
+1)^{2}r^{2}} \\ \label{heisa}
a& =\left[ C_{1}r^{\frac{8\kappa (\kappa +1)(2\kappa -1)}{4\kappa ^{2}+1}%
}+C_{2}r^{\frac{(2\kappa -1)\left( 8\kappa ^{2}+4\kappa +1\right) }{4\kappa
^{2}+1}}\right] ^{\frac{1}{2}} \\ \label{heisn}
n& =\frac{\sqrt{C_{1}}\left( 8\kappa ^{2}+2\kappa -1\right) }{\left( 4\kappa
^{2}+1\right) \left( C_{1}+C_{2}r^{\frac{-8\kappa ^{2}+6\kappa -1}{4\kappa
^{2}+1}}\right) ^{\frac{1}{2}}} \\ \label{heispsi}
\psi & =\frac{(2\kappa +1)(4\kappa -1)}{4\kappa ^{2}+1}\ln \left( \frac{r}{%
\beta }\right)  \\
b& =r,
\end{align}%
\end{subequations}
with $C_{1}$ and $C_{2}$ being constants of integration. The on mass shell
constant values of $I_{1}$ and $I_{2}$ are in this case
\begin{align}
I_{1}& =\frac{2^{3\kappa +2}C_{2}f_{0}\kappa (\kappa +1)\left( 2\kappa
^{2}+\kappa -1\right) ^{2\kappa }\left( 4\kappa ^{2}+1\right) }{\sqrt{C_{1}}%
(1-4\kappa )^{\kappa +1}(2\kappa +1)^{2\kappa +1}} \\
I_{2}& =\frac{2^{3\kappa +1}C_{2}f_{0}(\kappa +1)^{2\kappa +1}(2\kappa
-1)^{2\kappa }\left( 4\kappa ^{2}+1\right) }{\sqrt{C_{1}}(2\kappa
+1)^{2\kappa +1}(1-4\kappa )^{\kappa }}, 
\end{align}%
resulting in relation between them, i.e. $I_1 = \frac{2\kappa(2\kappa+1)}{1-4\kappa}\, I_2$. Upon setting $C_{2}=0$ we recover the solution of the $I_{1}=I_{2}=0$ case.

It is interesting to note that solution \eqref{heisQ}-\eqref{heispsi} also recovers the Schwarzschild solution when $\kappa=0$. It quite remarkable that the Schwarzschild metric is recovered as the limiting case of two distinct solutions of the theory. The special solution of this case with $I_1\neq0\neq I_2$ and the general solution \eqref{fsolb}-\eqref{fsoln} of the $I_1=0$ case. However, we can distinguish a subtle difference when approaching the new limits. For small values of $\kappa =\varepsilon $, $\varepsilon ^{2}\rightarrow 0$, the solution of this section becomes
\begin{align}
a\left( r\right)  =&\sqrt{\left( C_{1}+\frac{C_{2}}{r}\right) }-\frac{ (4 C_1 r+C_2)\ln r}{\sqrt{r} \sqrt{C_1 r+C_2}} \varepsilon +O\left( \varepsilon
^{2}\right) , \\
n \left( r\right)  =& - \sqrt{\frac{C_1 r}{C_2 + C_1 r}}+ \frac{\sqrt{C_1} \sqrt{r} (2 (C_1 r+C_2)+3 C_2 \ln r)}{(C_1 r+C_2)^{3/2}} \varepsilon +O\left( \varepsilon
^{2}\right) .
\end{align}
We remark that for large values of the radius there is a large deviation
from the Schwarzschild spacetime. Specifically, if we take the limit $r\rightarrow+\infty$ in the above expressions we will see that $a(r)$ goes to infinity and the Minkowski spacetime is not recovered. This is in contrast to what happens with solution \eqref{fsolb}-\eqref{fsoln}, where we have seen that the flat space limit was recovered when considering the same limit.

\section{Exact solutions for nonmetricity scalar-tensor theory}

We consider the case of nonmetricity scalar-tensor theory (\ref{sd.01}) in
the static spherically symmetric background (\ref{lineelgen}) for connection
(\ref{consol1}) with the constraints (\ref{set2}). The minisuperspace
Lagrangian for the field equations is%
\begin{equation} \label{Lagphi}
\hat{L}_{2}\left( n,a,\dot{a},b,\dot{b},\phi ,\dot{\phi},\psi ,\dot{\psi}%
\right) = e^\phi \left[ \frac{1}{n} \left( 2 b \dot{a} \dot{b} + a \dot{b}^2 + a b^2 \dot{\psi} \dot{\phi} - \frac{\omega }{2}  a b^2 \dot{\phi}^2 \right) + n a\left( 1 + \frac{\dot{\phi}}{\dot{\psi}} + b^2 \hat{V}(\phi) \right)\right],
\end{equation}
where $\hat{V}(\phi)=e^{-\phi}V(\phi)$. The Lagrangian produces Euler-Lagrange equations which are equivalent to the gravitational field equations and which are given by
\begin{equation} \label{eqphi1}
\frac{1}{n^2} \left( 2 b \dot{a} \dot{b} + a \dot{b}^2 + a b^2 \dot{\psi} \dot{\phi} - \frac{\omega }{2}  a b^2 \dot{\phi}^2 \right) - a\left( 1 + \frac{\dot{\phi}}{\dot{\psi}} + b^2 \hat{V}(\phi) \right) =0,
\end{equation}%
\begin{equation} \label{eqphi2}
-2\dot{b}^{2}+4b\dot{b}\left( \frac{\dot{n}}{n}-\dot{\phi}\right)
+ 2 n^{2}\left( 1- \hat{V}\left( \phi \right) b^{2}+ \frac{\dot{\phi}}{\dot{\psi%
}}\right) +b^{2}\left(2\dot{\phi}\dot{\psi}- \omega \dot{\phi}^{2}\right) + 4\dot{\phi}\left( 1 - b \dot{b}  \right) -4b \ddot{b} =0,
\end{equation}%
\begin{equation} \label{eqphi3}
\frac{a }{b^2}\left(\dot{b}^2-n^2 \left(\frac{\dot{\phi}}{\dot{\psi}}+1\right)\right) + 2 \dot{a} \left(\frac{\dot{n}}{n}-\frac{\dot{b}}{b} \right) + \dot{\phi} \left(a \dot{\psi}-2 \dot{a}\right) - \frac{\omega}{2} a \dot{\phi}^2 - a n^2 \hat{V}(\phi) -2 \ddot{a} =0 ,
\end{equation}
\begin{equation} \label{eqphi4}
  \frac{b^2 }{n}\left(\frac{a \dot{n} \dot{\phi}}{n}-\dot{a} \dot{\phi}-\frac{2 a \dot{b} \dot{\phi}}{b}-a \dot{\phi}^2\right) + \frac{n}{\dot{\psi}^2} \left(\dot{a} \dot{\phi} +\frac{a \dot{n} \dot{\phi}}{n} + a \dot{\phi}^2\right) + \frac{a }{n} \left(\frac{n^2}{\dot{\psi}^2}-b^2\right) \ddot{\phi} - \frac{2 a  n \dot{\phi}}{\dot{\psi}^3} \ddot{\psi} =0 ,
\end{equation}
\begin{equation} \label{eqphi5}
  \begin{split}
    & n^2 \left(a-\frac{\dot{a}}{\dot{\psi}}-a b^2 \left(\hat{V}'(\phi)+\hat{V}(\phi)\right)\right) + 2 b \dot{b} \left(\dot{a} + \frac{a \dot{b}}{2 b}+\omega  a \dot{\phi} \right) + \omega  b^2 \dot{\phi} \left( \dot{a}-\frac{a \dot{n}}{n}\right) - b \dot{\psi} \left(b \dot{a}+2 a  \dot{b} - \frac{a b  \dot{n}}{n}\right) \\
    & + \frac{1}{2} \omega  a  b ^2 \dot{\phi}^2 - \frac{a  n  \dot{n}}{\dot{\psi}} + a \left(\frac{n^2}{\dot{\psi}^2}-b^2\right)  \ddot{\psi} + \omega  a b ^2 \ddot{\phi}=0 .
  \end{split}
\end{equation}
The Lagrangian \eqref{Lagphi}, and subsequently the equations \eqref{eqphi1}-\eqref{eqphi5}, are transformed to \eqref{minlag2} and \eqref{euln}-\eqref{eulpsi} with the change
\begin{equation}
  V(\phi) = Q f'(Q)-f(Q), \quad \phi = \ln (2 f'(Q)) ,
\end{equation}
under the condition that there is no kinetic term for $\phi$, i.e. $\omega=0$. For the power law theory, $f(Q)=f_0 Q^{1+\kappa}$, which we examine here, the above expressions reduce to
\begin{equation}
   V(\phi) = V_0 e^{\frac{1+\kappa}{\kappa} \phi} , \quad Q= Q_0 e^{\frac{\phi}{\kappa}},
\end{equation}
and for the exact correspondence we have, $V_0= \frac{\kappa  f_0^{-1/\kappa }}{(2 (\kappa +1))^{\frac{\kappa +1}{\kappa }}}$ and $Q_0= \left(2 f_0 (1+\kappa)\right)^{-1/\kappa}$. 

Finally, the conserved quantities (\ref{cons1}) and (\ref{cons2}) in these ``coordinates'' are given (up to multiplicative constants) by
\begin{equation}
\hat{I}_{1}=\frac{a}{n}\left( b^{2}-\left( \frac{n}{\dot{\psi}}\right)
^{2}\right) e^{\phi }\dot{\phi}
\end{equation}%
\begin{equation}
\hat{I}_{2}= \frac{e^{\phi }}{n\dot{\psi}}\left( \kappa a\left(n^{2}+b\dot{\psi}\left( b \dot{\psi} -2\dot{b}\right) \right) - b^{2}\dot{a}\dot{\psi}\right) .
\end{equation}
All previously presented solutions are valid and satisfy the equations of such a theory with the spacetime metric and the nonmetricity being exactly the same. The difference is that instead of a $f(Q)\sim Q^{1+\kappa}$ function there is a potential $V(\phi) \sim e^{\frac{1+\kappa}{\kappa} \phi}$ of a scalar field without kinetic term, given by
\begin{equation}
  \phi = \kappa  \ln \left(2^{1/\kappa } (\kappa +1)^{1/\kappa } f_0^{1/\kappa } Q\right),
\end{equation}
where $Q$ in the above expression is substituted by \eqref{fsolQ}, \eqref{fsolQsp}, \eqref{fsolQ00} or \eqref{heisQ} depending on the solution we are using.

\section{Interpretation of the solutions}

The line elements that emerge from the above analysis do not obey the Einstein's field equations since they are produced with the use of modified symmetric teleparallel gravity. In order to assign a physical meaning to them, from the perspective of general relativity, we can proceed as follows:

First we calculate the Einstein tensor $G_{\alpha \beta}=R_{\alpha \beta}-\frac{1}{2}R g_{\alpha \beta}$ for the metric in question and interpret it as an effective energy-momentum tensor by writing $G_{\alpha \beta}=T^{(imf)}_{\alpha \beta}$.

We then check if this $T^{imf}_{\alpha \beta}$ fits to that of an imperfect fluid described by the energy-momentum tensor of the form
\bal
T^{(imf)}_{\alpha \beta}=\left( \rho+p \right)u_\alpha u_\beta+p g_{\alpha \beta}+2q_{(\alpha}u_{\beta)}+\pi_{\alpha \beta},
\eal
where $\rho$ is the energy density of the fluid, $u^\alpha$ the 4--velocity, $q^\alpha$ the heat flux vector, $p$
the pressure and $\pi_{\alpha \beta}$ the anisotropic stress tensor, satisfying $u^\mu q_\mu=0, \pi_{\mu\nu} u^\mu, \pi^\mu{}_\mu=0$.

In order to make the identification possible we define the tensor $\Pi_{\alpha \beta}=G_{\kappa \lambda} h^\kappa{}_\alpha h^\lambda{}_\beta$ and we use the relations
\bsub
\begin{alignat}{2}
 \rho &=G_{\alpha \beta}u^\alpha u^\beta,  &  \quad p&=\frac{1}{3}\Pi_\alpha{}^\alpha,\\
q_\kappa&=-G_{\alpha \beta}u^\alpha h^\beta{}_\kappa, & \quad \pi_{\alpha \beta}&=\Pi_{\alpha \beta}-p h_{\alpha \beta},
\end{alignat}
\esub
 where $h_{\alpha \beta}$ is the projection tensor orthogonal to velocity $u_\alpha$ defined by
\bal
h_{\alpha \beta} = g_{\alpha \beta}+ u_\alpha u_\beta \quad \text{with} \quad u_\alpha u^\alpha=-1,
\eal
for the calculational details see \cite{1985Madsen,Madsen:1988ph,Pimentel:1989bm}.
Furthermore the kinematical quantities of the fluid that are of interest and appear in the decomposition of the covariant derivative of the velocity \cite{ellis2012relativistic}
\bal
\nabla_\alpha u_\beta=-\dot{u}_\alpha u_\beta+\omega_{\alpha \beta}+\sigma_{\alpha \beta}+\frac{1}{3}\theta h_{\alpha \beta},
\eal
are
\bal
\dot{u}_\alpha=u^\kappa\nabla_\kappa u_\alpha, \quad \theta=\nabla_\alpha u^\alpha, \quad \sigma_{\alpha \beta}=\nabla_{(\alpha} u_{\beta)}+\dot{u}_{(\alpha}u_{\beta)}-\frac{1}{3}\theta h_{\alpha \beta}, \quad \omega_{\alpha \beta}=\nabla_{[\alpha} u_{\beta]}+\dot{u}_{[\alpha}u_{\beta]},
\eal
i.e. the acceleration, the expansion, the shear and the rotation of the fluid respectively.

In the following subsections we will apply the above technique to the solutions derived before. The appropriate 4-velocity for the metric \eqref{lineelgen} reads $u^\alpha=(a,0,0,0)$.

\subsection{Solution $I_1 \neq 0$ and $I_2 \neq 0$}

The energy fluid density $\rho$ and the pressure $p$ are equal to
\bal
\rho		& = \frac{4 \kappa  (\kappa +1) (2 \kappa -1) (6 \kappa +1)}{(2 \kappa+1)^2 (4 \kappa -1)^2} r^{-2}+
	\frac{C_2}{C_1} \frac{2  \kappa  (2 \kappa -3) \left(4 \kappa ^2+1\right) }{ (2 \kappa +1)^2 (4 \kappa -1)^2}
	\,r^{-\frac{16\kappa^2 - 6\kappa+3 }{4 \kappa ^2+1}}\\
p 		& =	\frac{4 (2 \kappa-1 )^2 \kappa  (\kappa +1) \left(8 \kappa ^2+12\kappa -1\right)}{3 (4 \kappa -1)^2 (2 \kappa +1)^2 } r^{-2}+ \frac{C_2}{C_1} \frac{4 \kappa  \left(4 \kappa ^2+1\right) \left(8 \kappa ^3-2
   \kappa +1\right)}{3 (2 \kappa +1)^2 (4 \kappa -1)^2} \, r^{-\frac{16 \kappa ^2-6 \kappa +3}{4 \kappa ^2+1}}
\eal

From the above we arrive at a non constant parameter $w$ in the equation of state $p=w\,\rho$. Interesting enough this parameter becomes constant
\bal
w=\frac{(2 \kappa -1) \left(8 \kappa ^2+12 \kappa -1\right)}{3(6 \kappa +1)}
\eal
when $C_2=0$.

The fluid has zero heat flux $q_\alpha = 0$ and suffers anisotropic stresses through $\pi_{\alpha\beta}$ which reads
\bal
\pi_{\alpha\beta} =
\begin{pmatrix}
0		&		0																	&		0													&		0 \\
0		&		\dfrac{\alpha_1 C_1}{C_2 r^{\beta_1}+ C_1 r^2} + \tilde{\gamma}_1\, r^{-2}	&		0													&		0 \\
0		&		0																	&		\alpha_2 + \beta_2\dfrac{C_2}{C_1}\, r^{\tilde{\gamma}_2}	&		0 \\
0		&		0																	&		0													&		\left(\alpha_2 + \beta_2\dfrac{C_2}{C_1}\, r^{\tilde{\gamma}_2}\right) \sin^2\theta
\end{pmatrix}
\eal
where
\bal
\alpha_1 &=-\frac{2 \kappa  (4 \kappa -1) (6 \kappa +1)}{3 \left(4\kappa ^2+1\right)}, &
\beta_1 &= \frac{6 \kappa +1}{4 \kappa ^2+1}, &
\tilde{\gamma}_1 &=-\frac{2 \kappa  \left(16 \kappa ^3-24 \kappa ^2-10 \kappa+5\right)}{3 \left(4 \kappa ^2+1\right)} \non \\
\alpha_2 &=\frac{4 \kappa  \left(4 \kappa ^2+1\right) \left(4 \kappa ^3-3 \kappa +1\right)}{3 (4\kappa-1)^2 (2\kappa+1)^2}, & \beta_2 &=-\frac{(4\kappa^2 +1)^2}{2(2\kappa+1)^2(4\kappa-1)^2}\, \tilde{\gamma}_1, &
\tilde{\gamma}_2 &=\frac{-8 \kappa ^2+6 \kappa -1}{4 \kappa ^2+1}\non
\eal

Furthermore there is no expansion $\theta$, no shear $\sigma_{\alpha\beta}$ and no rotation $\omega_{\alpha\beta}$ for this fluid.

\subsection{Solution $I_1 = 0$ and $I_2 \neq 0$ }

In this case the components of the energy momentum tensor $T^{imf}_{\alpha\beta}$ are quite complicate expressions, thus we present only the special case where $\beta = 0$, which turns to be quite interesting. With this assumption the energy fluid density $\rho$ and the pressure $p$ are equal to
\bal
\rho &= \frac{4 \kappa  (\kappa +1) (2 \kappa -1) (6 \kappa +1)}{(2 \kappa +1)^2 (4 \kappa -1)^2} \, r^{-\frac{2 \left(4 \kappa ^2+1\right)}{\alpha^2(2 \kappa +1) (4 \kappa -1)}}, \\
p &= \frac{4 \kappa  (\kappa +1) (2 \kappa -1)^2 \left(8 \kappa ^2+12 \kappa -1\right)}{3 \alpha^2 (2 \kappa +1)^2 (4 \kappa -1)^2} \, r^{-\frac{2 \left(4 \kappa ^2+1\right)}{(2 \kappa +1) (4 \kappa -1)}}
\eal
which yield to a constant parameter $w$
\bal
w=\frac{(2 \kappa -1) \left(8 \kappa ^2+12 \kappa -1\right)}{3 (6\kappa +1)}.
\eal

The fluid has zero heat flux $q_\alpha = 0$ and suffers anisotropic stresses through $\pi_{\alpha\beta}$ which reads
\bal
\pi_{\alpha\beta} =
\begin{pmatrix}
0		&		0						&		0		&		0 \\
0		&		-\dfrac{2\tilde{\alpha}}{r^2}		&		0		&		0 \\
0		&		0						&		\tilde{\alpha}	&		0 \\
0		&		0						&		0		&		\tilde{\alpha} \sin^2\theta
\end{pmatrix}
\eal
where
\bal
\tilde{\alpha}=\dfrac{4 \kappa  (\kappa +1) (2 \kappa -1)^2 \left(4 \kappa ^2+1\right)}{3 (2 \kappa +1)^2 (4 \kappa -1)^2}
\eal

Also in this case there is no expansion $\theta$, no shear $\sigma_{\alpha\beta}$ and no rotation $\omega_{\alpha\beta}$ for the fluid in question, irrespectively of the value of $\beta$.

As we previously discussed, $\beta$ is associated to the mass of the Schwarzschild solution when $\kappa=0$. Although we cannot claim that $\beta$ acquires the same physical meaning for the generic $\kappa\neq 0$ case, we could interpret the above expressions as a zero-mass limit of the obtained solution. As is well-known, a zero mass for the Schwarzschild case ($\kappa=0$) leads to the Minkowski spacetime. However, for the modified theory ($\kappa\neq0$), the metric is not Minkowski as $\beta=0$, it is equivalent to a metric resulting from a perfect fluid with a linear equation of state. Thus, we see that the theory even in pure vacuum, in the sense of not having some point mass placed at zero radius, leads to non-trivial metrics.

\subsection{Solution $I_1 = 0$ and $I_2 \neq 0$ on special case $\kappa=\frac{1}{4}$}

The energy fluid density $\rho$ and the pressure $p$ are equal to
\bal
\rho &= -\frac{b_2}{b_1^2}\frac{\left(3 b_2+4 r^{1/3}\right) e^{\frac{6 b_2}{r^{1/3}}}}{3 r^{8/3}}, \\
p &= \frac{b_2^2}{b_1^2}\frac{e^{\frac{6 b_2}{r^{1/3}}} \left(54 b_2 r^{1/3}+18 b_2^2+35 r^{2/3}\right)}{3 r^{8/3} \left(6 b_2+5 r^{1/3}\right){}^2}
\eal
which yield to a parameter $w$
\bal
w=-\frac{b_2 \left(54 b_2 r^{1/3}+18 b_2^2+35 r^{2/3}\right)}{\left(3 b_2+4 r^{1/3}\right) \left(6 b_2+5 r^{1/3}\right){}^2}.
\eal

The fluid has zero heat flux $q_\alpha = 0$ and suffers anisotropic stresses through $\pi_{\alpha\beta}$ which reads
\bal
\pi_{\alpha\beta} =
\begin{pmatrix}
0		&		0						&		0		&		0 \\
0		&		-\dfrac{2\tilde{\alpha}}{r^2}		&		0		&		0 \\
0		&		0						&		\tilde{\alpha}	&		0 \\
0		&		0						&		0		&		\tilde{\alpha} \sin^2\theta
\end{pmatrix}
\eal
where this time
\bal
\tilde{\alpha}=\frac{b_2^2 \left(18 b_2 r^{1/3}+9 b_2^2+10 r^{2/3}\right)}{3 r^{2/3} \left(6 b_2+5 r^{1/3}\right){}^2}
\eal

Also and in this case there is no expansion $\theta$, no shear $\sigma_{\alpha\beta}$ and no rotation $\omega_{\alpha\beta}$ for the fluid in question.

\subsection{Solution $I_1 = 0$ and $I_2 = 0$}

The energy fluid density $\rho$ and the pressure $p$ are equal to
\bal
\rho &= \frac{4 \kappa  (\kappa +1) (2 \kappa -1) (6 \kappa +1)}{(2 \kappa +1)^2 (4 \kappa -1)^2 r^2}, \\
p &= \frac{4 \kappa  (\kappa +1) (2 \kappa -1)^2 \left(8 \kappa ^2+12 \kappa -1\right)}{3 (2 \kappa +1)^2 (4 \kappa -1)^2 r^2}
\eal
which yield to a constant parameter $w$
\bal
w=\frac{(2 \kappa -1) \left(8 \kappa ^2+12 \kappa -1\right)}{3 (6 \kappa +1)}.
\eal

The fluid has zero heat flux $q_\alpha = 0$ and suffers anisotropic stresses through $\pi_{\alpha\beta}$ which reads
\bal
\pi_{\alpha\beta} =
\begin{pmatrix}
0		&		0					&		0		&		0 \\
0		&		-\dfrac{\tilde{\beta}}{r^2}	&		0		&		0 \\
0		&		0					&		\tilde{\alpha}	&		0 \\
0		&		0					&		0		&		\tilde{\alpha} \sin^2\theta
\end{pmatrix}
\eal
where
\bal
\tilde{\alpha}=\frac{4 \kappa  (\kappa +1) (2 \kappa -1)^2 \left(4 \kappa ^2+1\right)}{3 (2 \kappa +1)^2 (4 \kappa -1)^2}, \quad
\tilde{\beta}=\frac{8 (1-2 \kappa )^2 \kappa  (\kappa +1)}{12 \kappa ^2+3}.
\eal

Also and in this case there is no expansion $\theta$, no shear $\sigma_{\alpha\beta}$ and no rotation $\omega_{\alpha\beta}$ for the fluid in question.

\section{Conclusions}

STEGR and its extensions/modifications have gathered significant attention because they open new directions for describing physical phenomena. While the separation of the connection from the metric tensor in STEGR leads to general relativity (GR), this is not the case for its extensions. In these extended theories, the connection plays a crucial role in the gravitational model, introducing new degrees of freedom that stem from the definition of the connection. The minimum number of dynamical degrees of freedom is introduced when the connection is defined in the coincidence gauge, but this definition is not unique.

Theories of this type have also gathered criticism based on the difficulty of directly observing effects related to nonmetricity \cite{Golovnev}. This is a valid point, especially for the case of STEGR, which is dynamically equivalent to GR. The modifications of the theory however, like $f(Q)$-gravity, propagate additional degrees of freedom - more from other modified theories like $f(R)$-gravity. This can be seen from the Hamiltonian formalism of the theory, although there is some controversy in the literature about how many degrees of freedom there are in total \cite{DBfQ1,DBfQ2,DBfQ3}. In principle there should be some way of tracing these extra physical degrees of freedom and their possible effects. In addition, there have been proposed mechanisms related to possible observational effects owed to nonmetricity \cite{expnonm1,expnonm2,expnonm3}. These mostly depend on the way matter may couple to nonmetricity and it is an intriguing subject, see \cite{Delhom} for a study on what constitutes a minimal coupling in theories with nonmetricity and torsion.

In this work we were preoccupied with the case of a static, spherically symmetric spacetime within the framework of $f(Q)$ gravity, to recover the GR limit of STEGR, the connection must be defined in the non-coincidence gauge. Two scalar fields are introduced by the theory. One accounts for the higher-order derivatives of the nonlinear function $f(Q)$, while the second scalar field governs the evolution of the connection. A similar dynamical behavior is present in the nonmetricity scalar-tensor theory.

The field equations in these two gravitational theories possess the property of admitting a minisuperspace description. This allowed us to apply the method of variational symmetries to construct conservation laws for specific functional forms of the free functions in these theories. The construction of conservation laws was essential in this study, as it allowed us to reduce the order of the field equations and derive closed-form analytic solutions. The geometric and physical properties of these solutions were studied. The resulting spacetimes are consistent with the theorem presented in \cite{Baha1} stating that for nonlinear $f(Q)$ theories $g_{tt}$ is different from $g_{\bar{r}\bar{r}}^{-1}$ when $\bar{r}$ is the radial distance. 

What we have not discussed yet are the algebraic properties of the derived spacetimes. We investigated the isometries of the spacetimes and, as expected, we found that all solutions admit the three Killing vector fields describing the spherical symmetry of the metric, i.e.
\bal\label{sph_killing}
Y_1 = \cos\phi \,\partial_\theta - \cot\theta \sin\phi \,\partial_\phi, \,
Y_2 = \sin\phi \,\partial_\theta + \cot\theta \cos\phi \,\partial_\phi, \,
Y_3=\partial_\phi
\eal
along with the $Y_4 = \partial_t$ which describes the static character of the metric. However, the solution with $I_1=I_2=0$, besides the above fields, also admits a homothetic Killing vector $H$, i.e. $\mathcal{L}_H g_{\alpha\beta} = g_{\alpha\beta} $, of the form
\bal\label{homo}
H=\frac{-8\kappa^3+4\kappa+1}{2(4\kappa^2+1)}\,t \,\partial_t + r\,\partial_r.
\eal

It is quite interesting that the same theory $f(Q)\sim Q^{1+\kappa}$, supports two distinct solutions \eqref{oursol} and \eqref{Heissol}, which both have as a limiting case the Schwarzschild solution of GR. The first is the general solution of $I_1=0$, $I_2\neq 0$ and the second a special solution of $I_1 \neq 0 \neq I_2$. The metric \eqref{oursol} leads to an asymptotically Minkowski spacetime as $\kappa\rightarrow 0$, while the \eqref{Heissol} doesn't. However, we cannot say whether \eqref{oursol} can be mapped into black hole solutions in the coordinates where $b$ becomes the radial variable. Such coordinate transformations, will in principle involve regions where they become complex and this may affect the geometric properties in ways  that we cannot predict. Complex transformations have been used in the literature to map different geometries and also connect different black hole solutions \cite{BHcpx1,BHcpx2,BHcpx3}.

Another point we want to refer to is that the static spherically symmetric solutions we report here, can also be  ``transferred'' to a cosmological setting. To do this one needs to interchange $t$ and $r$ so that the static spherically symmetric line element we consider becomes a Kantowski-Sachs spacetime. The conservation laws we derived here are hence also true for this cosmological counterpart of a power law theory $f(Q)\sim Q^{1+\kappa}$. Lastly, we need to mention that, from the solutions we present here, we excluded from our analysis the case $f(Q)\sim \sqrt{Q}$, which leads to $Q=0$.

In the future, we plan to extend this analysis by introducing a matter source and studying the effects of conformal transformations on the physical properties of the solutions. We will also search for the general solution of the model presented here and further investigate the possible existence of black hole solutions.

\begin{acknowledgments}
AP thanks the support of VRIDT through Resoluci\'{o}n VRIDT No. 096/2022 and Resoluci\'{o}n VRIDT No. 098/2022. Part of this study was supported by FONDECYT 1240514.
\end{acknowledgments}

\end{document}